\documentclass[12pt,preprint]{aastex}
\newcommand{\beq}{\begin{equation}}
 \newcommand{\eeq}{\end{equation}}
 \newcommand{\beqn}{\begin{eqnarray}}
 \newcommand{\eeqn}{\end{eqnarray}}

\begin{document}
 
 \title{\bf{A deep study of an intermediate age open cluster SAI 35 ({\bf Juchert 20}) using ground based imaging and Gaia~EDR3 astrometry.}}
 
 \author{D. Bisht$^1$, Qingfeng Zhu$^1$, R. K. S. Yadav$^2$, Geeta Rangwal$^3$, Alok Durgapal$^3$, Devesh P. Sariya$^4$ and Ing-Guey Jiang$^4$} 
 
 \affil{
 {$^1$Key Laboratory for Researches in Galaxies and Cosmology, University of Science and Technology of China, Chinese Academy of Sciences,
      Hefei, Anhui, 230026, China}\\
      {email: dbisht@ustc.edu.cn}\\
 {$^2$Aryabhatta Research Institute of Observational Sciences,Manora Peak, Nainital 263 002, India}\\
 {$^3$Center of Advanced Study, Department of Physics, D. S. B. Campus, Kumaun University Nainital 263002, India}\\ 
 {$^4$Department of Physics and Institute of Astronomy, National Tsing-Hua University, Hsin-Chu, Taiwan}\\ 
}
 
\begin{abstract}

We present CCD $UBVI$ photometric study of poorly studied intermediate age open cluster SAI 35 (Juchert 20) for the first time.
To accomplish this study, we also used LAMOST~DR5, 2MASS, and Gaia~EDR3 databases. We identified 214 most probable 
cluster members with membership probability higher than $50\%$. 
The mean proper motion of the cluster is found as $\mu_{\alpha}cos\delta=1.10\pm0.01$ and
$\mu_{\delta}=-1.66\pm0.01$ mas yr$^{-1}$. We find the normal interstellar extinction law using the various two-color diagrams. The age, 
distance, reddening, and radial velocity of the cluster are estimated to be $360\pm40$ Myr, $2.9\pm0.15$ kpc, $0.72\pm0.05$ mag and 
$-91.62\pm6.39$ km/sec.
The overall mass function slope for main-sequence stars is found to be $1.49\pm0.16$ within the mass range 1.1$-$3.1 $M_\odot$, which
is in agreement with Salpeter's value within uncertainty. The present study demonstrates that SAI 35 is a dynamically relaxed.
Galactic orbital parameters are determined using Galactic potential models. We found that this object follows a circular path
around the Galactic center.

\end{abstract}
 
 
\section{Introduction}
\label{Intro}

\begin{figure*}
\begin{center}
\includegraphics[width=8.5cm, height=8.5cm]{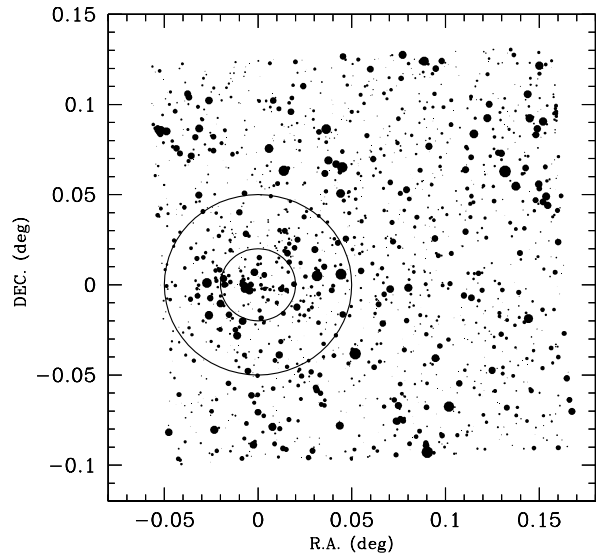}
\caption{Finding chart of the stars in the field of SAI 35. Filled circles of different sizes represent brightness of
the stars. Smallest size denotes stars of $V\sim$20 mag. Open outer circle represent the cluster size and inner circle
represent core region.}
\label{id}
\end{center}
\end{figure*}

Open clusters (OCs) have long been used for the understanding of stellar physics, to study the star formation scenario and the overall
structure of the Milky Way (MW). The open cluster SAI 35 (Juchert 20) ($\alpha_{2000} = 04^{h}10^{m}47.0^{s}$,
$\delta_{2000}=46^{\circ} 52^{\prime} 01^{\prime\prime}$; $l$=154$^\circ$.494, $b$=-3$^\circ$.422) is located in the
second Galactic quadrant. This object is listed in the catalog http://ocl.sai.msu.ru/catalog/.

This open cluster is listed in the catalog given by Kronberger et al. (2006). Kharchenko et al. (2012, 2013) cataloged
the proper motions, distance, reddening, and log(age) value of SAI 35 as (-2.36, -6.11) mas/yr, 2812 pc, 0.791 mag, and 8.22, respectively
based on 2MASS and PPMXL catalog. Dias et al. (2014) derived proper motion values of this object as -0.11 and -0.90 mas/yr based on UCAC4
catalog. The integrated $JHK_{s}$ magnitudes and luminosity function has been estimated by Kharchenko et al. (2016). A catalog of cluster
membership has been given by Sampedro et al. (2017) based on UCAC4 data. Cantat-Gaudin et al. (2018) has made a catalog for cluster members and
obtained fundamental parameters of SAI 35 based on Gaia DR2 data.

OCs generally suffer from the field star contamination towards the fainter ends in the main sequence. Hence, knowledge about the cluster
membership status of stars is necessary. The Gaia DR2 catalog was made public on 25$^{th}$ April 2018
(Gaia Collaboration et al. 2018a,b). The (early) Third Gaia Data Release (hereafter EDR3; Gaia Collaboration et al. 2020)
was made public on 3$^{rd}$ December 2020. EDR3 consists the central coordinates, proper motions in right ascension and declination
and parallax angles $(\alpha, \delta, \mu_{\alpha}cos\delta, \mu_{\delta}, \pi)$ for around 1.46 billion sources with a limiting
magnitude of 3 to 21 mag in $G$ band. The Gaia EDR3 data is much precise and accurate in comparison to the second data release. The
Gaia data can be used for the precise estimation of membership of cluster members to have a better understanding of the
fundamental parameters of OCs. Cantat-Gaudin et al. (2018) obtained membership probabilities for 1229 OCs using Gaia DR2 but it is
limited to source brighter than 18 mag in $G$ band. Recently they added many clusters to get a catalog of 1481 OCs (Cantat-Gaudin \& Anders 2020).
In this paper, we have estimated membership probabilities of stars towards the region of SAI 35 down to $\sim$20 mag in $V$ band. 

\begin{figure}
\begin{center}
\includegraphics[width=8.5cm, height=8.5cm]{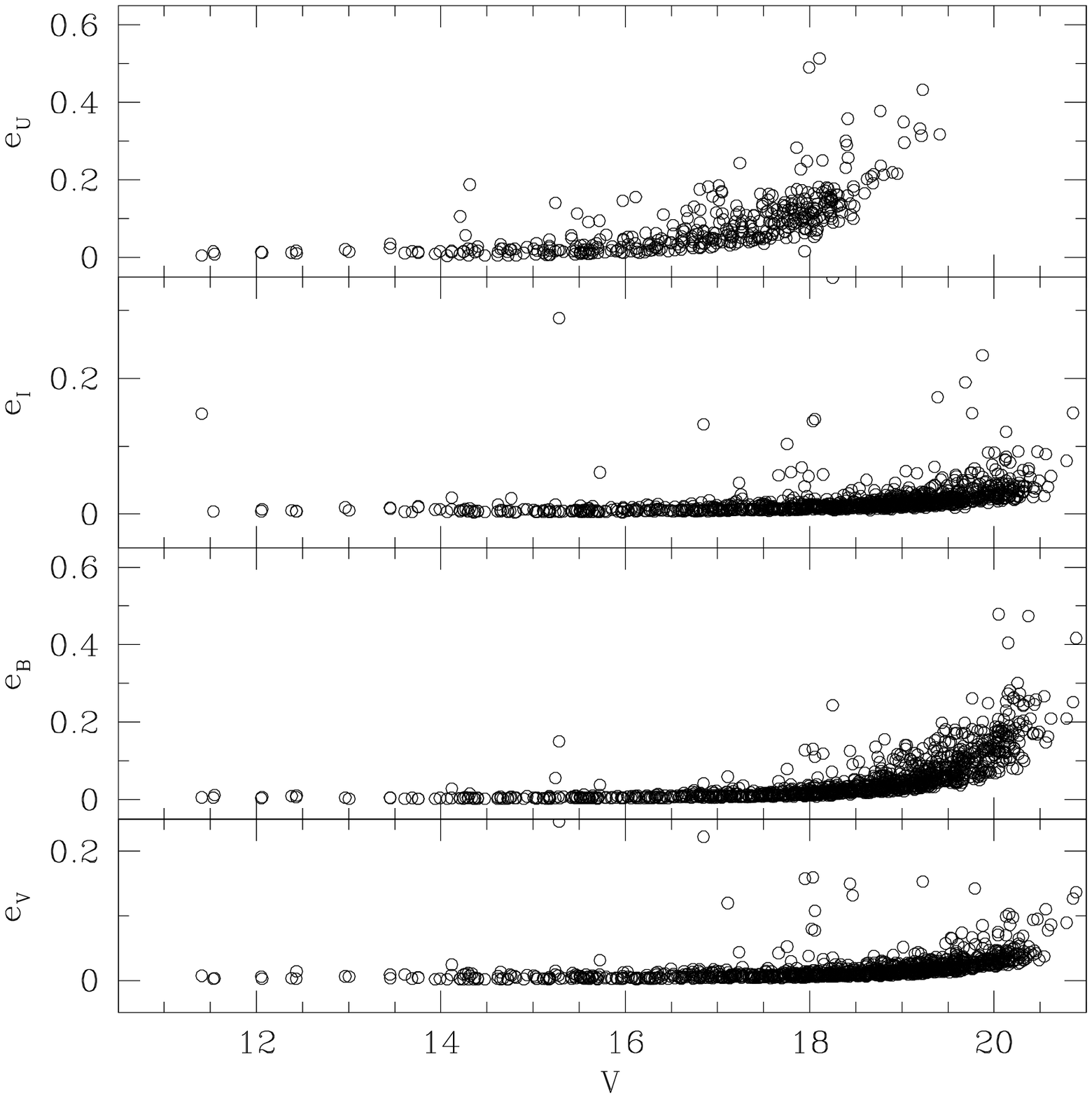}
\caption{Photometric errors in different bands against $V$ magnitude.}
\label{error_ubvi}
\end{center}
\end{figure}

Spectroscopic data in Gaia EDR3 is the same as of Gaia DR2. Soubiran et al. (2018) reported 6D phase space information of 861 stars
clusters using weighted mean radial velocity based on Gaia DR2. More than $50\%$ of the above cataloged clusters have radial velocity
information using less than 3 stars. The cluster studied here (SAI 35) is not listed in the above spectroscopic catalog given by
Soubiran et al. (2018). We have taken radial velocity data from the fifth release (DR5) of the Large sky Area Multi-Object
fiber Spectroscopic Telescope (LAMOST, Cui et al. 2012; Zhao et al. 2012; Luo et al. 2012).

In recent years, many authors have estimated the present day mass function for a sample of clusters (e.g., Hur et al. 2012,
Khalaj \& Baumgardt 2013, Dib et al. 2017, Joshi et al. 2020). It is still a debatable question to all researchers whether
the initial mass function (IMF) of OCs is universal in time and space or it depends on the star forming conditions (Elmegreen 2000,
Kroupa 2002, Bastian et al. 2010; Dib \& Basu 2018). OCs are considered as crucial objects to investigate the dynamical evolution
of the stellar system (Bisht et al. 2019). The study of mass-segregation in OCs give a clue about the distribution of low and high
mass stars towards the cluster region.

The paper is organized as follows. Section 2 describes the observations, data reduction procedure and other used data sets. The completeness
of CCD data is described in Section 3. Section 4 deals with the study of proper motion and determination of the membership probability of
stars. The structural properties and derivation of fundamental parameters using the most probable cluster members have been carried out
in Sections 5 and 6. The dynamical study of the cluster is discussed in Section 7. 
The cluster's orbit is studied in Section 8. 
Finally, the conclusions are presented in Section 9.

\begin{table}
\begin{center}
\caption{Log of observations, with dates and exposure times for each passband.}
\vspace{0.5cm}
\begin{tabular}{ccc}
\hline\hline
Band  &Exposure Time &Date\\
&(in seconds)   & \\
\hline\hline
$U$&1200$\times$2, 300$\times$1&21$^{st}$ January 2010 \\
$B$&900$\times$2, 240$\times$1&,,\\
$V$&600$\times$3, 180$\times$2&,,\\
$I$&300$\times$2, 60$\times$2&,,\\
\hline
\end{tabular}
\label{log}
\end{center}
\end{table}

\section{Observations and data analysis}

\begin{figure}
\begin{center}
\includegraphics[width=8.0cm, height=8.0cm]{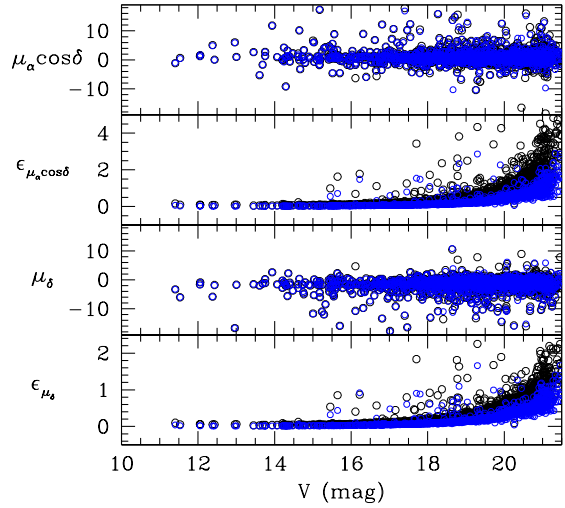}
\caption{Plot of proper motions and their errors versus $V$ magnitude. Blue circles are Gaia EDR3 sources while the black filled circles show the Gaia DR2 sources towards the area of SAI 35.}
\label{error_pm}
\end{center}
\end{figure}
We carried out CCD $UBVI$ observations of stars in the region of open cluster SAI 35 on $21^{st}$ January 2010. We used
a 104-cm Sampurnanand reflector telescope (f/13) located at Aryabhatta Research Institute of Observational Sciences,
Manora Peak, Nainital, India. Images were acquired using a 2K$\times$2K CCD which has 24 $\mu$m square  pixel size,
resulting in a scale of 0$^{\prime\prime}$.36 pixel$^{-1}$ and a square field of view of 12.$^{\prime}$6 size. The CCD
gain was 10 e$^{-}$/ADU while the readout noise was 5.3 e$^{-}$. In order to improve the S/N ratio, the observations
were taken in the 2$\times$2 pixel$^2$ binned mode. Table \ref{log} lists the date of observations together with the
filters used and the corresponding range of exposure time. The identification map observed by us is shown in Fig \ref{id}.
\begin{figure}
\begin{center}
\includegraphics[width=6.5cm, height=6.5cm]{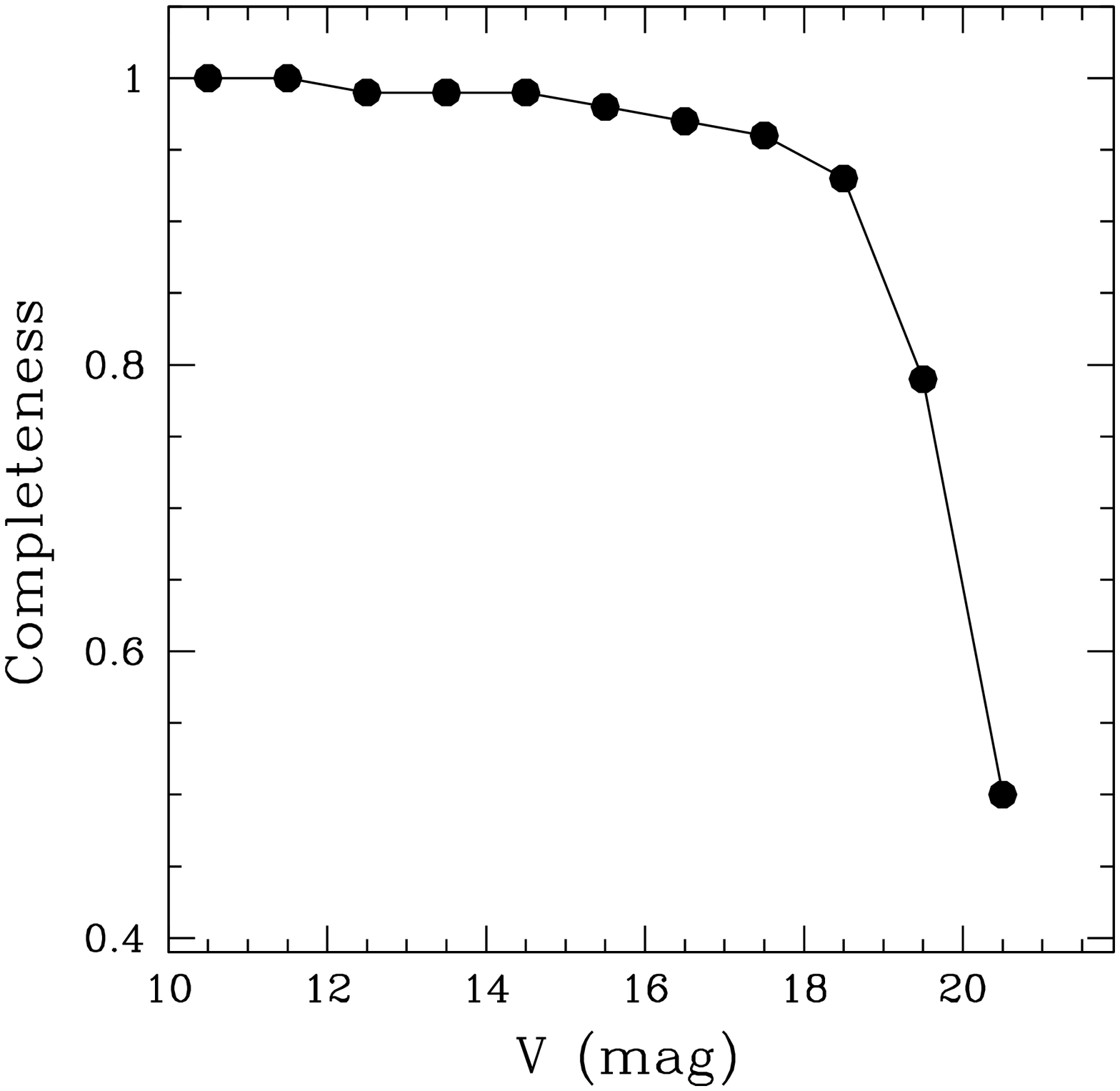}
\vspace{-0.5cm}
\caption{Variation of completeness factor versus $V$ magnitude.}
\label{comp_fig}
\end{center}
\end{figure}

Several biases and twilight flat-field images were taken in $UBVI$ filters, during the observing night. IRAF\footnote{ IRAF
is distributed by the National Optical Astronomical Observatory which are operated by the Association of Universities for Research
in Astronomy, under contract with the National Science Foundation} data reduction package has been used for the initial processing
of raw photometric data which consists of bias subtraction, flat fielding and cosmic ray removal. We used DAOPHOT software to
estimate the stellar magnitudes. The instrumental magnitudes were derived through point spread function (PSF) fitting using
DAOPHOT/ALLSTAR (Stetson 1987, 1992) package. To estimate the PSF, we used several well isolated stars for the entire frame.
The Gaussian function was used as an analytical model PSF. The shape of the PSF was made to vary quadratically with the position on
the image. Appropriate aperture corrections were determined by using isolated and unsaturated bright stars in the image.

We have cross-identify the stars of different frames and filters using the DAOMATCH/DAOMASTER programme available in
DAOPHOT II. To determine the transformation coefficients from instrumental to standard magnitudes, CCDLIB, CCDSTD routines
have been used. Finally, standard magnitudes and colors of all the stars have been obtained using the routine FINAL.

\begin{table}
\begin{center}
\caption{Derived Standardization coefficients and its errors.}
\vspace{0.5cm}
\begin{tabular}{ccc}
\hline\hline
Filter  &   Colour Coeff. $(C)$ & Zeropoint $(Z)$\\
\hline\hline
&SAI 35&\\ 

$U$&$-0.01\pm$0.02&$7.80\pm$0.01\\
$B$&$-0.02\pm$0.01&$5.44\pm$0.01\\
$V$&$-0.04\pm$0.01&$5.12\pm$0.01\\
$I$&$-0.05\pm$0.02&$5.53\pm$0.02\\
\hline
\end{tabular}
\label{c_coff}
\end{center}
\end{table}

\begin{table}
\centering
\caption{The rms global photometric errors as a function of $V$ magnitude.}
\vspace{0.5cm}
\begin{tabular}{ccccc}
\hline
$V$&$\sigma_{V}$&$\sigma_{B}$&$\sigma_{I}$&$\sigma_{U}$ \\
\hline
$10-11$&$0.04$&$0.05$&$0.04$&$0.05$ \\
$11-12$&$0.04$&$0.05$&$0.03$&$0.06$ \\
$12-13$&$0.05$&$0.05$&$0.05$&$0.07$ \\
$13-14$&$0.05$&$0.05$&$0.04$&$0.07$ \\
$14-15$&$0.05$&$0.06$&$0.05$&$0.08$ \\
$15-16$&$0.05$&$0.05$&$0.05$&$0.09$ \\
$16-17$&$0.06$&$0.05$&$0.06$&$0.12$ \\
$17-18$&$0.06$&$0.07$&$0.07$&$0.14$ \\
$18-19$&$0.07$&$0.07$&$0.07$&$0.19$ \\
$19-20$&$0.08$&$0.08$&$0.09$&$0.25$ \\
\hline
\end{tabular}
\label{g_error}
\end{table}

\subsection {Photometric calibration}

We have observed the standard field SA 98 (Landolt 1992) for SAI 35 during the observing night for photometric
calibration of CCD system. The 19 standard stars (SA98-650, 670, 653, 666, 671, 675, 676, 682, 685, 688, 1082, 1087,
1102, 1112, 724, 733, 1124, 1119, 1122) used in the calibrations have brightness and color range 9.54 $\le V \le$ 15.01
and $-0.004 < (B-V) < 1.909$ respectively. For the extinction coefficients, we assumed the typical values for the
ARIES observational site (Kumar et al. 2000). For translating the instrumental magnitude to the standard magnitude, the calibration
equations derived using the least square linear regression are as follows:

\begin{center}
   $u=U+Z_{U}+C_{U}(U-B)+k_{U}X$\\

   $b=B+Z_{B}+C_{B}(B-V)+k_{B}X$\\

   $v=V+Z_{V}+C_{V}(B-V)+k_{V}X$\\

   $i=I+Z_{I}+C_{I}(V-I)+k_{I}X$
\end{center}

\begin{figure}
\begin{center}
\includegraphics[width=8.5cm, height=10.5cm]{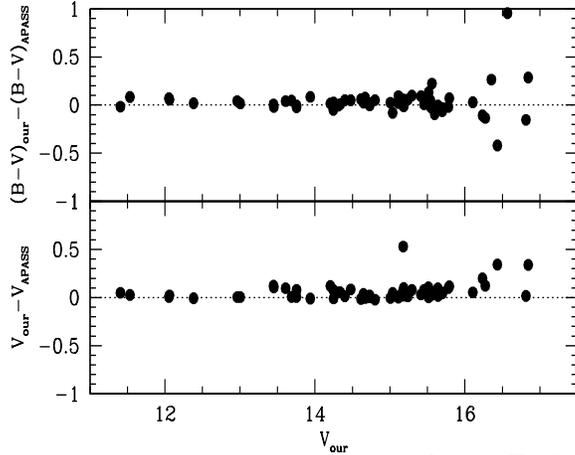}
\vspace{-3cm}
\caption{Differences between measurements presented in APASS catalog and in this study for V magnitude and (B-V) colors. Zero
difference is indicated by the dashed line.}
\label{match_p}
\end{center}
\end{figure}
where $u, b, v$ and $i$ denote the aperture instrumental magnitudes and $U, B, V$ and $I$ are the standard magnitudes whereas
airmass is denoted by $X$. The color coefficients (C) and zeropoints (Z) for different filters are listed in Table \ref{c_coff}.
The errors in zero points and color coefficients are $\sim$ 0.01-0.02 mag. The internal errors derived from DAOPHOT are
plotted against $V$ magnitude in Fig. \ref{error_ubvi}. This figure shows that the average photometric error is
$\le$ 0.06 mag for $B$, $V$ and $I$ filter at $V\sim19^{th}$ mag, while it is $\le$ 0.1 mag for $U$ filter at $V\sim18^{th}$
mag. Photometric global (DAOPHOT+Calibrations) errors are also calculated, which are listed in Table \ref{g_error}. For $V$
filter, the error are 0.06 at $V\sim$17 mag and 0.08 at $V\sim$20 mag. 

To transform the CCD pixel coordinates to celestial coordinates, we used the online digitized European Southern Observatory
catalog included in the $SKYCAT$ software as an absolute astrometric reference frame. The $CCMAP$ and $CCTRAN$ routine in
$IRAF$ was used to find a transformation equation which gives the celestial coordinates as a function of the pixel coordinates.
The resulting celestial coordinates have standard deviations of $\sim$0.1 arcsec in both right ascension and declination.

\subsection{Comparison with previous photometry}

\begin{table}
\centering
\caption{The difference in $V$ and $(B-V)$ between APASS catalog and present study. The standard
deviation in the difference for each magnitude bin is also given in the parentheses}
\vspace{0.5cm}
\begin{tabular}{ccc}
\hline
$V$&$\Delta{V}$&$\Delta{B-V}$ \\
\hline
$11-12$&$ 0.03~ (0.03)$&$0.02~ (0.03)$ \\
$12-13$&$ 0.02~ (0.05)$&$0.02~ (0.04)$ \\
$13-14$&$ 0.06~ (0.08)$&$0.03~ (0.10)$ \\
$14-15$&$ 0.07~ (0.08)$&$0.05~ (0.12)$ \\
$15-16$&$ 0.08~ (0.10)$&$0.04~ (0.18)$ \\
$16-17$&$ 0.09~ (0.14)$&$0.07~ (0.21)$ \\
\hline
\end{tabular}
\label{match_error}
\end{table}

To compare the photometry, We cross-matched the present catalog with
APASS catalog. For this matching the maximum difference in the positions of stars
is 1 arcsec. In this manner, we have found 68 common stars between these two 
catalogs. A comparison of $V$ magnitudes and $(B-V)$ color between the two catalogs is plotted against $V$ 
magnitude and shown in Fig \ref{match_p}.
The mean difference and standard deviation in per magnitude bin are listed in Table \ref{match_error}. The 
difference indicates that present $V$ and $(B-V)$ measurements are in fair agreement with those given in APASS catalog.

\subsection{Archived data}

\begin{table}
\large
\caption{Variation of completeness factor (CF) in the $V$, $(V-I)$
      diagram with the MS brightness.}
\begin{tabular}{cc}
\hline
$V$ mag range& CF\\
\hline
10-  11&1.00\\
11-  12&1.00\\
12-  13&0.99\\
13 - 14&0.99\\
14 - 15&0.99\\
15 - 16&0.98\\
16 - 17&0.97\\
17 - 18&0.96\\
18 - 19&0.93\\
19 - 20&0.79\\
20 - 21&0.50\\
\hline
\end{tabular}
\label{cf}
\end{table}

\subsubsection{Gaia EDR3}
Gaia~EDR3 (Gaia Collaboration et al. 2020) the database is used for the astrometric study of SAI 35. This data consist of positions
on the sky $(\alpha, \delta)$, parallaxes and proper motions ($\mu_{\alpha} cos\delta , \mu_{\delta}$) with a  limiting magnitude
of $G=21$ mag. The uncertainties in parallax values are $\sim$ 0.02-0.03 milliarcsecond (mas) for sources at $G\le15$ mag and
$\sim$ 0.07 mas for sources with $G\sim17$ mag. The proper motions with their respective errors are plotted against $G$ magnitude
in the right panel of Fig.~\ref{error_pm}. The uncertainties in the corresponding proper motion components are
$\sim$ 0.01-0.02 mas $yr^{-1}$ (for $G\le15$ mag), $\sim$0.05 mas $yr^{-1}$ (for $G\sim17$ mag),
$\sim$0.4 mas $yr^{-1}$ (for $G\sim20$ mag) and $\sim$1.4 mas $yr^{-1}$ (for $G\sim21$ mag). We have compared Gaia EDR3 proper
motion and their errors with Gaia DR2 data. We can clearly see in Fig \ref{error_pm}
that Gaia EDR3 data base is more precise than Gaia DR2. 

\begin{figure*}
\begin{center}
\includegraphics[width=8.5cm, height=8.5cm]{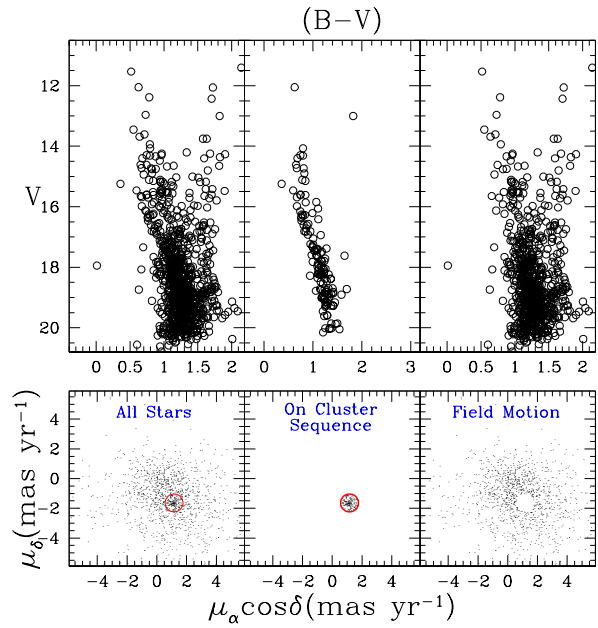}
\caption{(Bottom panels) Proper-motion vector point diagrams (VPDs) for SAI 35. (Top panels) $V$ versus $(B-V)$
color magnitude diagrams. (Left panel) The entire sample. (Center) Stars within the circle of 0.6~mas~yr$^{-1}$
radius centered around the mean proper motion of the cluster. (Right) Probable background/foreground field stars
in the direction of this object.}
\label{vpd}
\end{center}
\end{figure*}
\begin{figure*}
\begin{center}
\centering
\hbox{
\includegraphics[width=7.2cm, height=7.2cm]{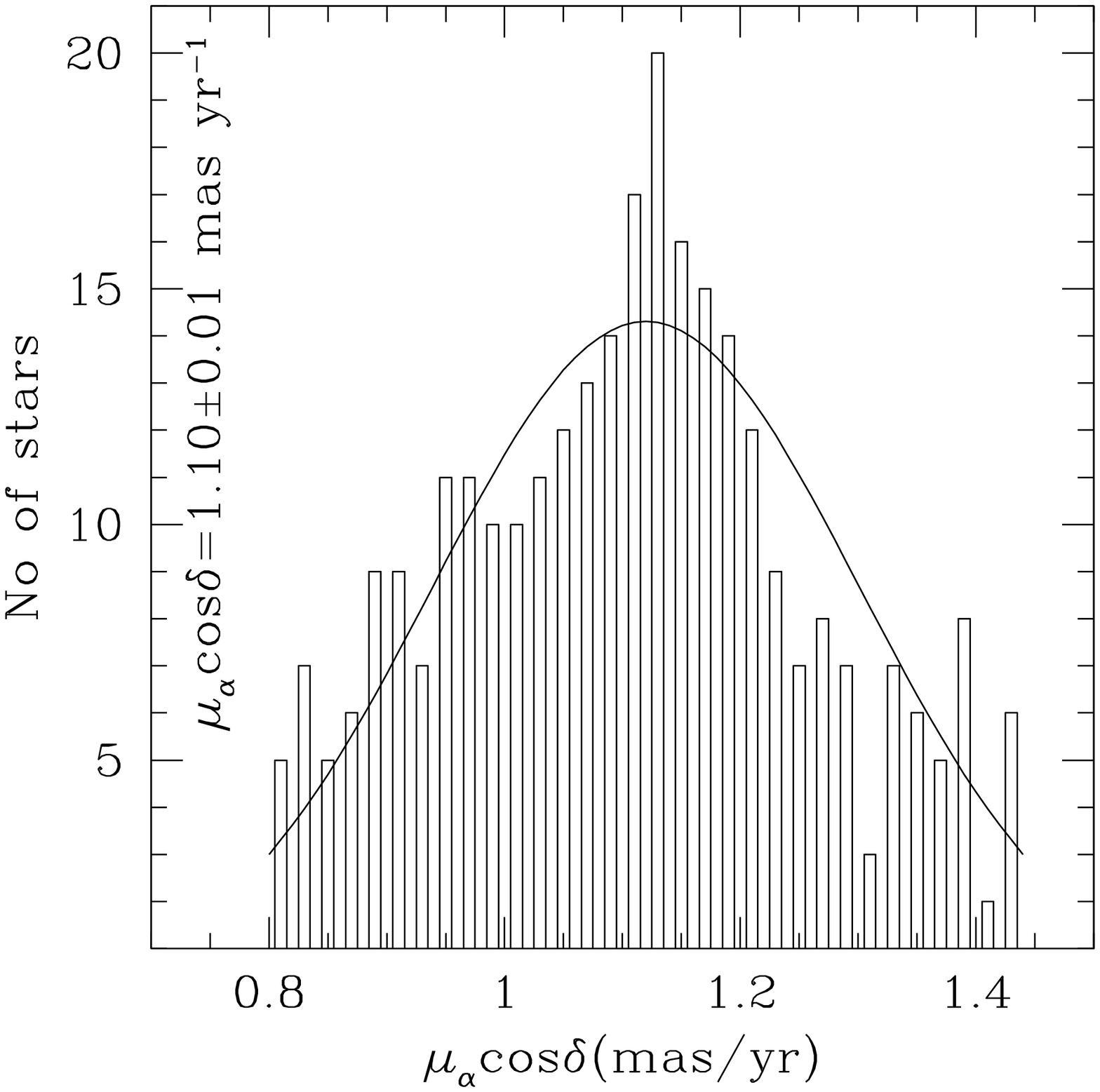}
\includegraphics[width=7.2cm, height=7.2cm]{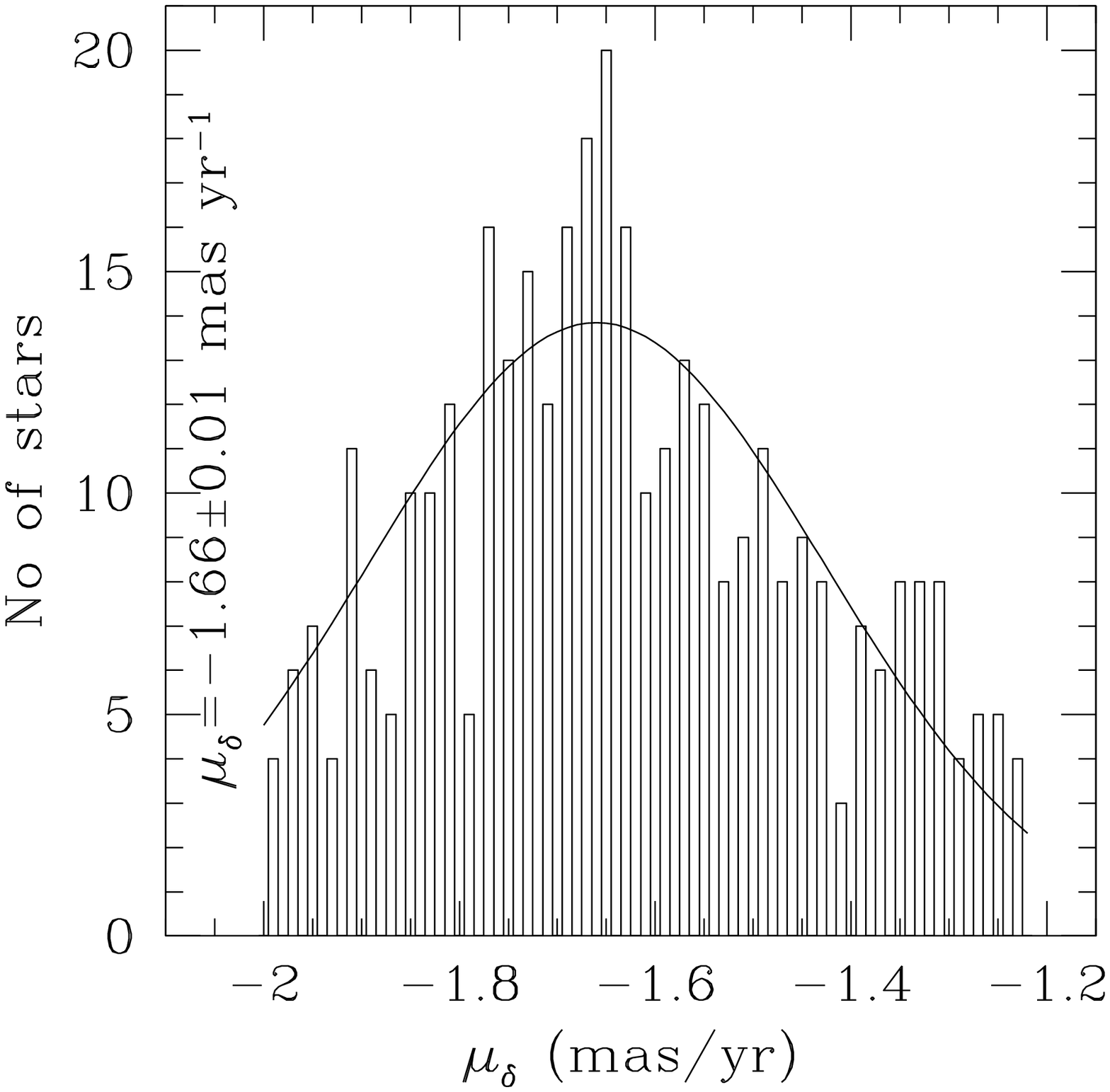}
}
\caption{Proper motion histograms of 0.1 mas~yr$^{-1}$ bins in right ascension and declination of the cluster SAI 35.
The Gaussian function fit to the central bins provides the mean values in both directions as shown in each panel.
}
\label{pm_dist1}
\end{center}
\end{figure*}
\subsubsection{2MASS data}

The near-Infrared $JHK$ photometric data for SAI 35 was taken from the Two Micron All-sky Survey (2MASS). 2MASS
consistently scanned the whole sky in three near-IR bands $J (1.25\micron)$, $H (1.65\micron)$ and $K (2.17\micron)$.
The 2MASS (Skrutskie et al. 2006) used two highly automated 1.3m aperture, open tube, equatorial fork-mount telescopes
(one at Mt. Hopkins, Arizona (AZ), USA and other at CTIO, Chile) with a 3-channel camera $(256\times256)$ array of
HgCdTe detectors in each channel). The $2MASS$ database contains photometry in the near infrared $J$, $H$ and $K$
bands to a limiting magnitude of 15.8, 15.1 and 14.3 respectively, with a signal to noise ratio (S/N) greater than 10.

\subsection{\bf APASS}

The American Association of Variable Star Observers (AAVSO) Photometric All-Sky Survey (APASS) is cataloged
in five filters: B, V (Landolt) and $g^{\prime}$, $r^{\prime}$, $i^{\prime}$, with $V$ band magnitude range
from 7 to 17 mag (Heden \& Munari 2014). The DR9 catalog covers about $99\%$ of the sky (Heden et al. 2016).
To compare the photometry, we have used data in $B$ and $V$ bands for SAI 35.

\subsubsection{LAMOST DR5}

LAMOST provided 9 million spectra with radial velocities in its fifth data release (DR5). This data also contains 5.3 million
spectra with stellar atmospheric parameters (effective temperature, surface gravity and metallicity). We used this data to obtain
the value of mean radial velocity and metallicity towards the region of SAI 35. The mean value of radial velocity has been
used to obtain the orbital parameters of the cluster.

\section{Completeness of the CCD data}

\begin{figure*}
\begin{center}
\hbox{
\includegraphics[width=7.5cm, height=7.5cm]{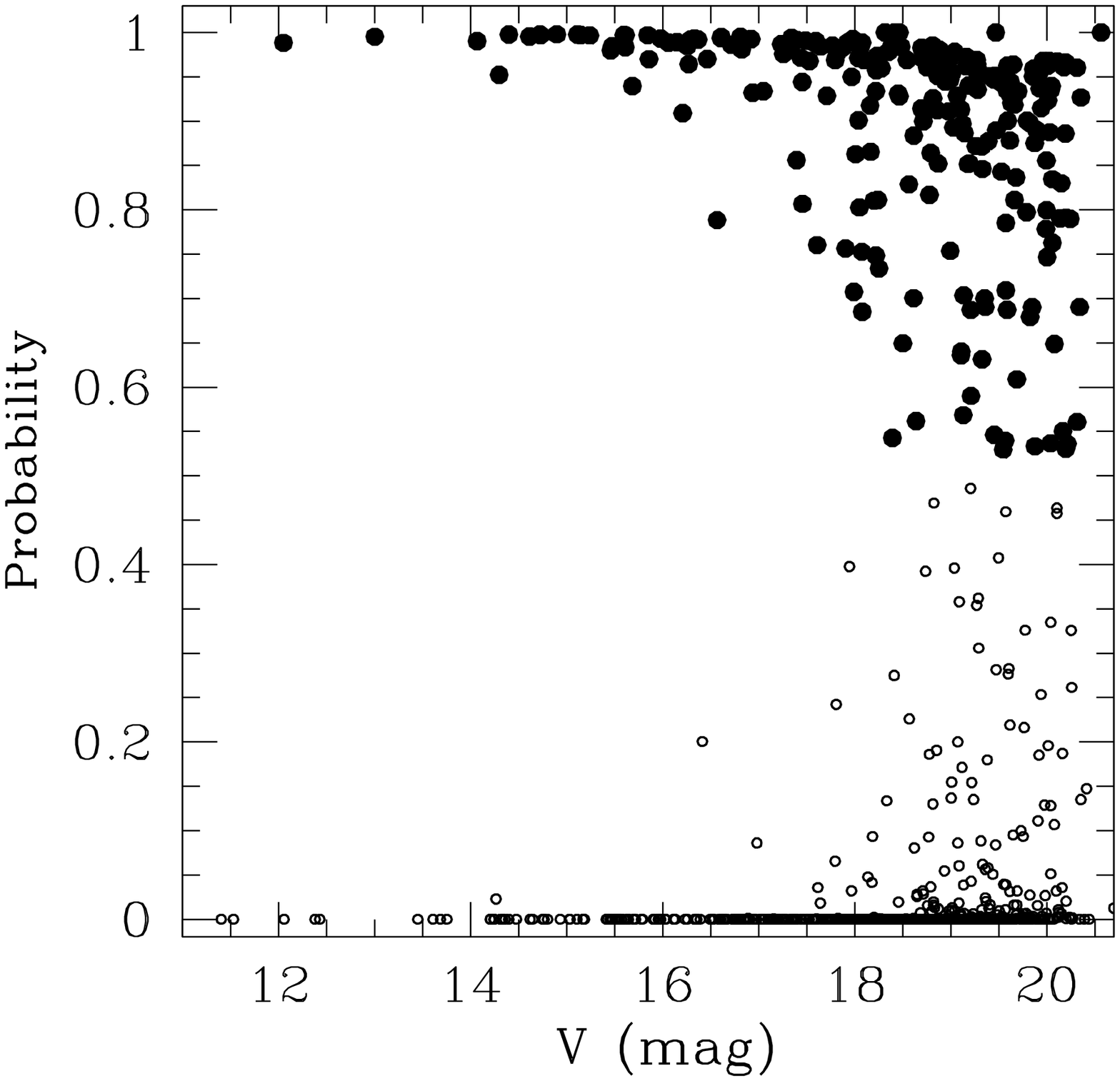}
\includegraphics[width=7.5cm, height=7.5cm]{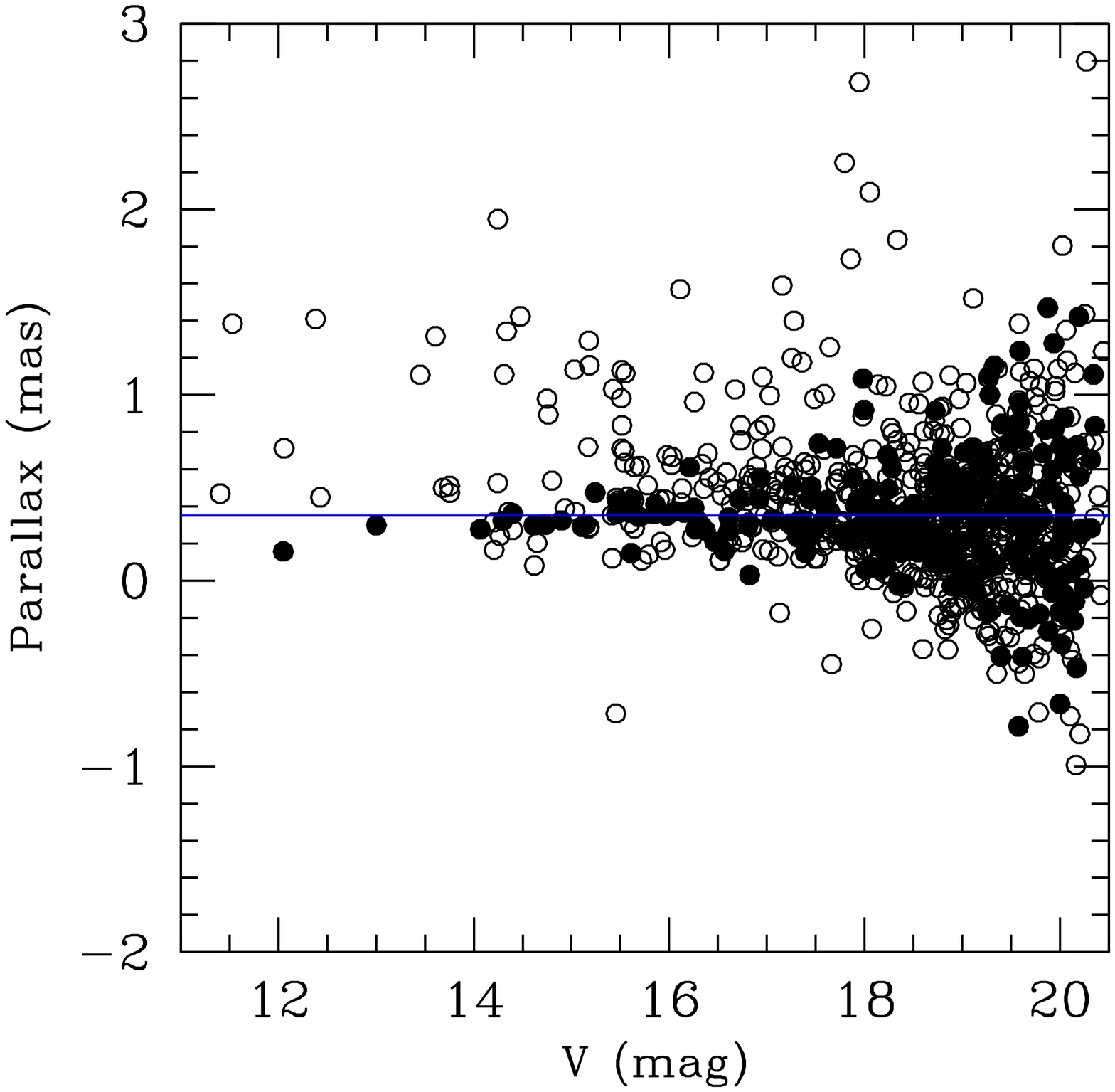}
}
\caption{(Left panel) Membership probability as a function of $V$ magnitude. (Right panel) Parallax as a function of
$V$ magnitude. The filled circles show the cluster members with membership probability higher than 50$\%$ in both the panels.}
\label{members}
\end{center}
\end{figure*}

\begin{figure}
\begin{center}
\includegraphics[width=7.5cm, height=7.5cm]{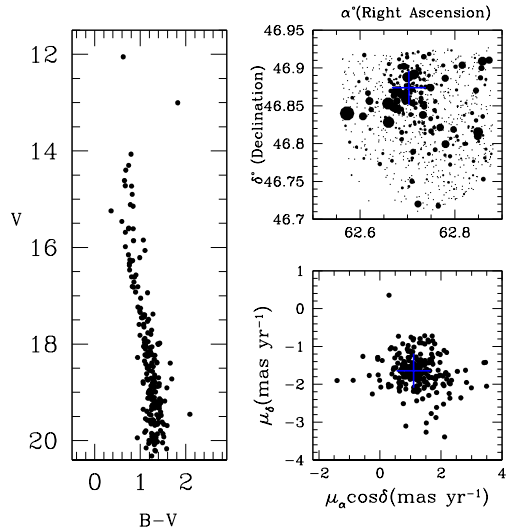}
\vspace{-0.5cm}\caption{($V, B-V$) CMD, identification chart and proper motion distribution of member stars with membership
probability higher than $50\%$. The plus sign indicates the cluster center in position and proper motions.}
\label{members1}
\end{center}
\end{figure}

\begin{figure}
\begin{center}
\includegraphics[width=6.5cm, height=6.5cm]{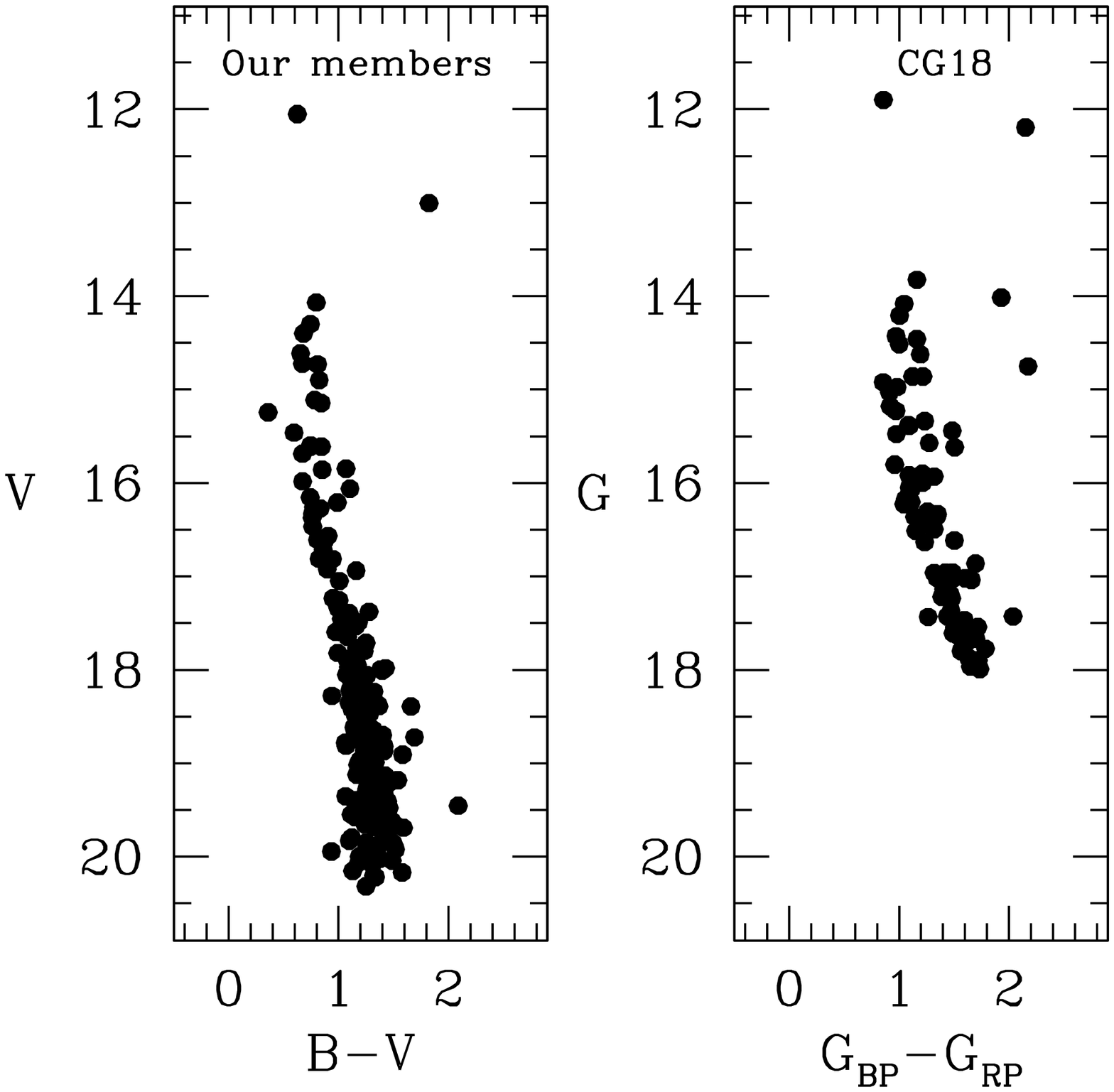}
\caption{The (V, $B-V$) CMD of our observed data and (G, $G_{BP}-G_{RP}$) CMD of Cantat-Gaudin et al. 2018 (CG18) catalog.
              All stars are plotted with membership probability higher than 50$\%$.}
\label{cmd_comp}
\end{center}
\end{figure}

The observational data may be incomplete because of the stellar crowding, saturation of bright stars, poor observing conditions,
the inefficiency of CCD data reduction programmes etc. The completeness correction is mandatory to compute luminosity
function of the stars in the cluster. To calculate the completeness level in our photometry for SAI 35, we performed the
artificial star (AS) test. We have randomly added only 10 to $15\%$ of actually detected stars into the original images so that the
crowding characteristics of the original images remain unchanged. The ADDSTAR routine in DAOPHOT II was used to determine
the completeness factor (CF). Detailed information about this experiment is given by Yadav \& Sagar (2002) and Sagar \& Griffiths (1998).
In the present analysis, we have adopted the method given by Sagar \& Griffiths (1998). Artificial stars with known magnitude and
position were added in the original $V$ frames. These images are re-reduced using a similar method that was adopted for the
original images. The ratio of recovered to added stars in different magnitude bins gives the CF. The CF derived in this way are
listed in Table \ref{cf} for SAI 35. Fig \ref{comp_fig} shows the variation of completeness factor versus $V$ magnitude. The
value of CF is found as $\sim$ 93$\%$ at V=19 mag.

\section{Proper motions and Field star separation}

Proper motion is a key parameter to separate field stars from the cluster region to truly understand the
main sequence of clusters. PM components ($\mu_{\alpha} cos{\delta}$, $\mu{\delta}$) are plotted as VPD in the bottom
panels of Fig.~\ref{vpd} after matching our observed $UBVI$ data with Gaia EDR3. The panels of top rows show the
corresponding $V$ versus ($B-V$) color-magnitude diagrams (CMDs). The left panel shows all the detected stars towards the region of
SAI 35, while the middle and right panels show the probable cluster members and field region stars. A circle of 0.6 mas
$yr^{-1}$ around the distribution of cluster stars in the VPD characterize our membership criteria. The chosen radius
is a compromise between losing member stars with poor PMs and contamination of field region stars (Sariya et al. 2015, Bisht et al. 2020).
We have also used mean parallax for the cluster member selection. We estimated the weighted mean of parallax for stars
inside the circle of VPD having $V$ mag brighter than 20$^{th}$ mag. We obtained the mean value of parallax as
$0.35\pm0.02$ mas. We considered a star as the most probable members if it lies within 0.6 mas yr$^{-1}$ radius in
VPD and has a parallax within 3$\sigma$ from the mean parallax of SAI 35. The CMD of the most probable cluster members
is shown in the upper-middle panel in Fig.~\ref{vpd}. In this figure, the main sequence of the cluster is identified.

To estimate the mean proper motion, we considered probable cluster members selected from VPD and CMD as shown in
Fig.~\ref{vpd}. By fitting the Gaussian function into the constructed histograms, provides mean proper motion in both
the directions of right ascension and declination as shown in Fig. \ref{pm_dist1}. In this way, we found the mean-proper
motion of SAI 35 as $1.10\pm0.01$ and $-1.66\pm0.01$ mas yr$^{-1}$ in $\mu_{\alpha} cos{\delta}$ and $\mu_{\delta}$
respectively. The estimated value of mean PM for this object is in very good agreement
with Cantat-Gaudin et al. (2018). Our derived values of mean proper motion is much reliable than Kharchenko et al. (2012, 2013) 
and Dias et al. (2014) because present estimated is based on accurate Gaia EDR3 proper motion data.


\subsection{Membership probabilities}
\label{MP}

\begin{figure*}
\begin{center}
\hbox{
\includegraphics[width=6.0cm, height=6.0cm]{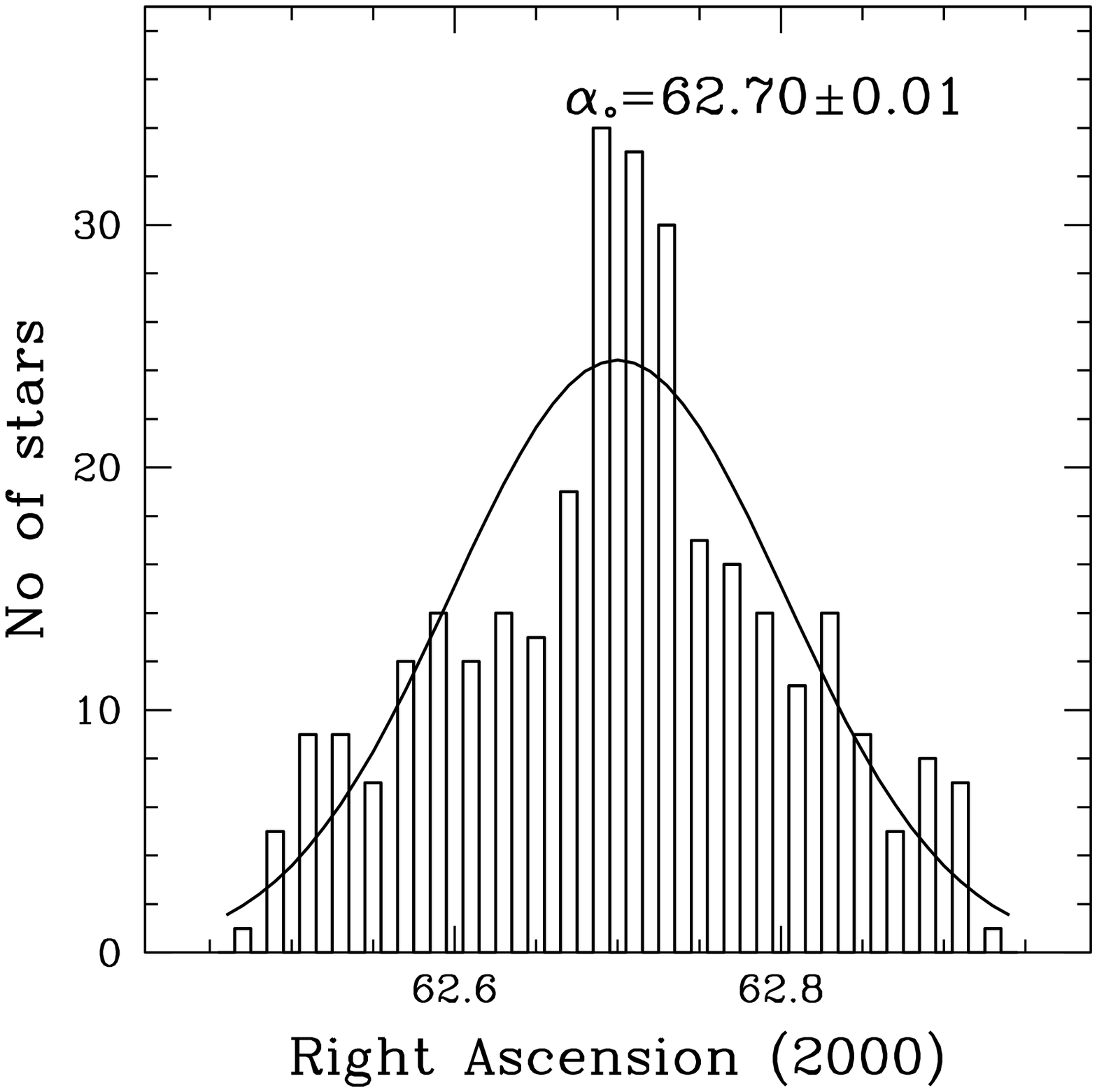}
\includegraphics[width=6.0cm, height=6.0cm]{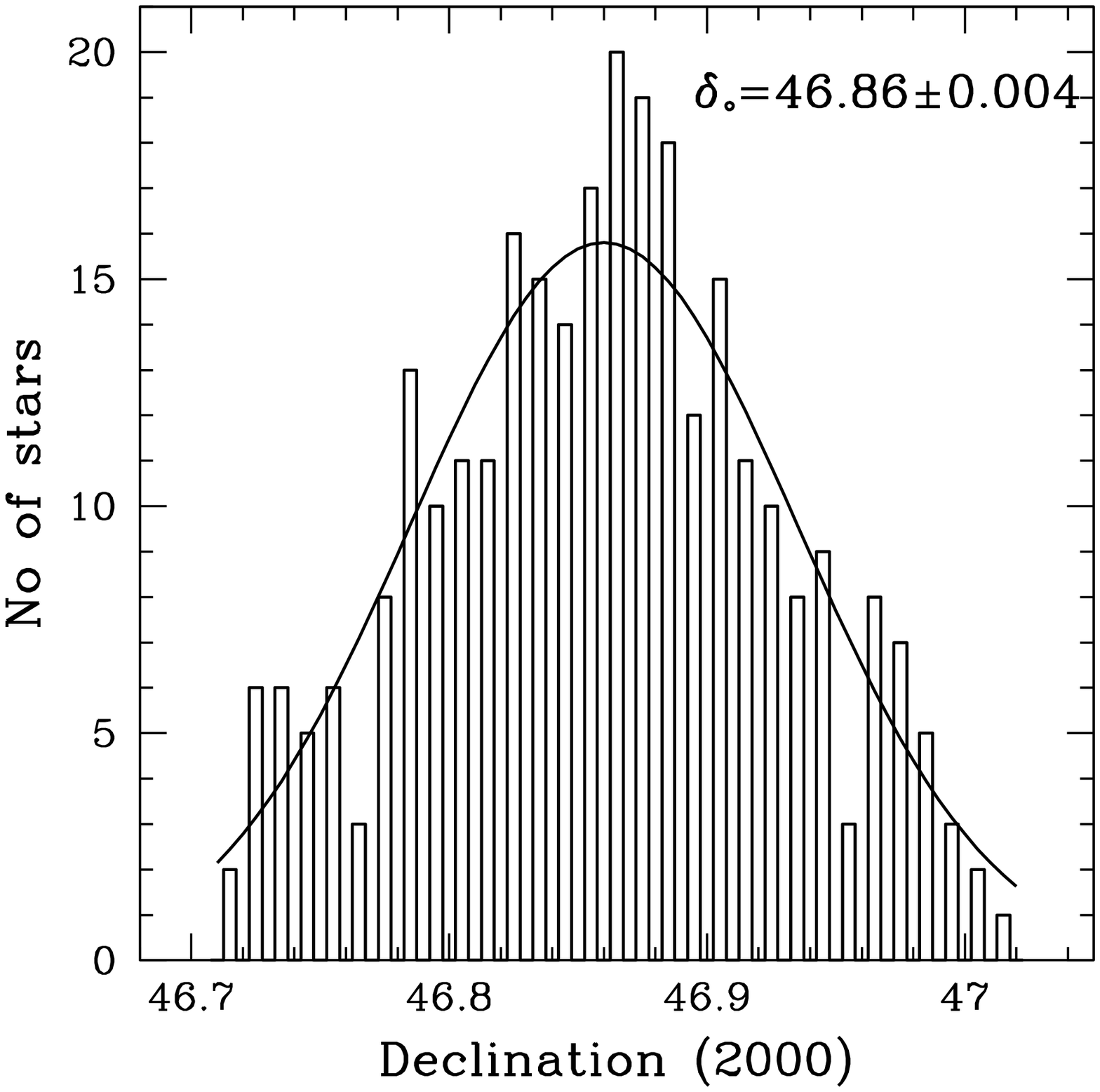}
}
\vspace{-0.5cm}\caption{Profiles of stellar counts across SAI 35 region. The Gaussian fits have been applied. The center of symmetry
about the peaks of Right Ascension and Declination is taken to be the position of the cluster's center.}
\label{center}
\end{center}
\end{figure*}

To estimate the membership probabilities of stars towards the region of SAI 35, we adopted the approach given by
Balaguer-N\'{u}\~{n}ez et al. (1998) by using proper motion and parallax data from Gaia EDR3. This membership
probability method has been used and described by various authors for a number of clusters (Kaur et al. 2020; Yadav et al. 2013;
Sariya et al. 2017; 2018; Bisht et al. 2020b).

To describe the distribution functions for cluster and field stars in the adopted method, we used only good stars which have PM
errors better than $\sim$0.5 mas~yr$^{-1}$. A clear crowding of stars can be seen at $\mu_{xc}$=1.10 mas~yr$^{-1}$,
$\mu_{yc}$=$-$1.66 mas~yr$^{-1}$ and in the circular region having radii of 0.6 mas~yr$^{-1}$. We estimated dispersion ($\sigma_c$)
in PMs as 0.09 mas~yr$^{-1}$ by fixing cluster distance as 2.9 kpc and the radial velocity dispersion of 1 km $s^{-1}$ for open star
clusters (Girard et al. 1989). For non members, we have estimated ($\mu_{xf}$, $\mu_{yf}$) = ($-$1.2, 0.2) mas yr$^{-1}$ and
($\sigma_{xf}$, $\sigma_{yf}$) = (2.9, 3.3) mas yr$^{-1}$.\\

We obtained 214 stars as cluster members after applying completeness to the observational CCD data along with with membership probability higher
than  50$\%$ and $V\le20$ mag. In the left panel of Fig.~\ref{members}, we plotted membership probability versus $V$ magnitude.
In this figure, cluster members and field stars are separated. In the right panel of this figure, we plotted $V$ magnitude versus
parallax of stars. In Fig.~\ref{members1}, we plotted identification chart, proper motion distribution and $V$ versus $B-V$ color-magnitude
diagram using the most probable cluster members. The most probable cluster members with high membership probability
$(\ge50\%)$ are shown by filled circles in Fig.~\ref{members} and Fig.~\ref{members1}.

Membership probability has been estimated by Cantat-Gaudin et al. (2018) up to 18.0 mag using the GAIA DR2 catalog for this cluster. To 
compare the membership probability, we plotted CMDs using our membership catalog and Cantat-Gaudin et al. (2018) catalog
as shown in Fig \ref{cmd_comp}. We used only probable stars with membership probability higher than 50$\%$. The CMDs plotted using
selected members of the cluster based on membership probability seem clean. Therefore, both the membership probabilities are comparable.

Blue Straggler stars (BSS) are intriguing objects in diverse environments such as OCs (Johnson \& Sandage 1955; Sandge 1962;
Ahumada \& Lapasset 1995). According to Sandage (1953), in CMDs of OCs BSS are found along the extension of the main sequence as the brighter
objects than the main sequence turn off points in the CMDs of clusters. BSS are considered crucial objects to study the interplay between
stellar evolution and stellar dynamics (Bailyn 1995). In this paper, we found one BSS which is located at a radial distance of $\sim$ 2.2
arcmin from the cluster's center. Our analysis strongly suggest that the BSS is a confirmed cluster member with a membership probability of
99$\%$.

\section{Structural analysis: radial density profile}

To estimate the center coordinates towards the area of SAI 35, we used the star-count method. The resulting histograms in
both the RA and DEC directions are shown in Fig.~\ref{center}. The Gaussian curve-fitting provides the central coordinates
as $\alpha = 62.70\pm0.01$ deg ($4^{h} 10^{m} 48^{s}$) and $\delta = 46.86\pm0.004$ deg ($46^{\circ} 51^{\prime} 36^{\prime\prime}$).
Our obtained values are very close to the values given by Sampedro et al. (2017) and Cantat-Gaudin et al. (2018).

\begin{figure}
\begin{center}
\includegraphics[width=7.0cm, height=7.0cm]{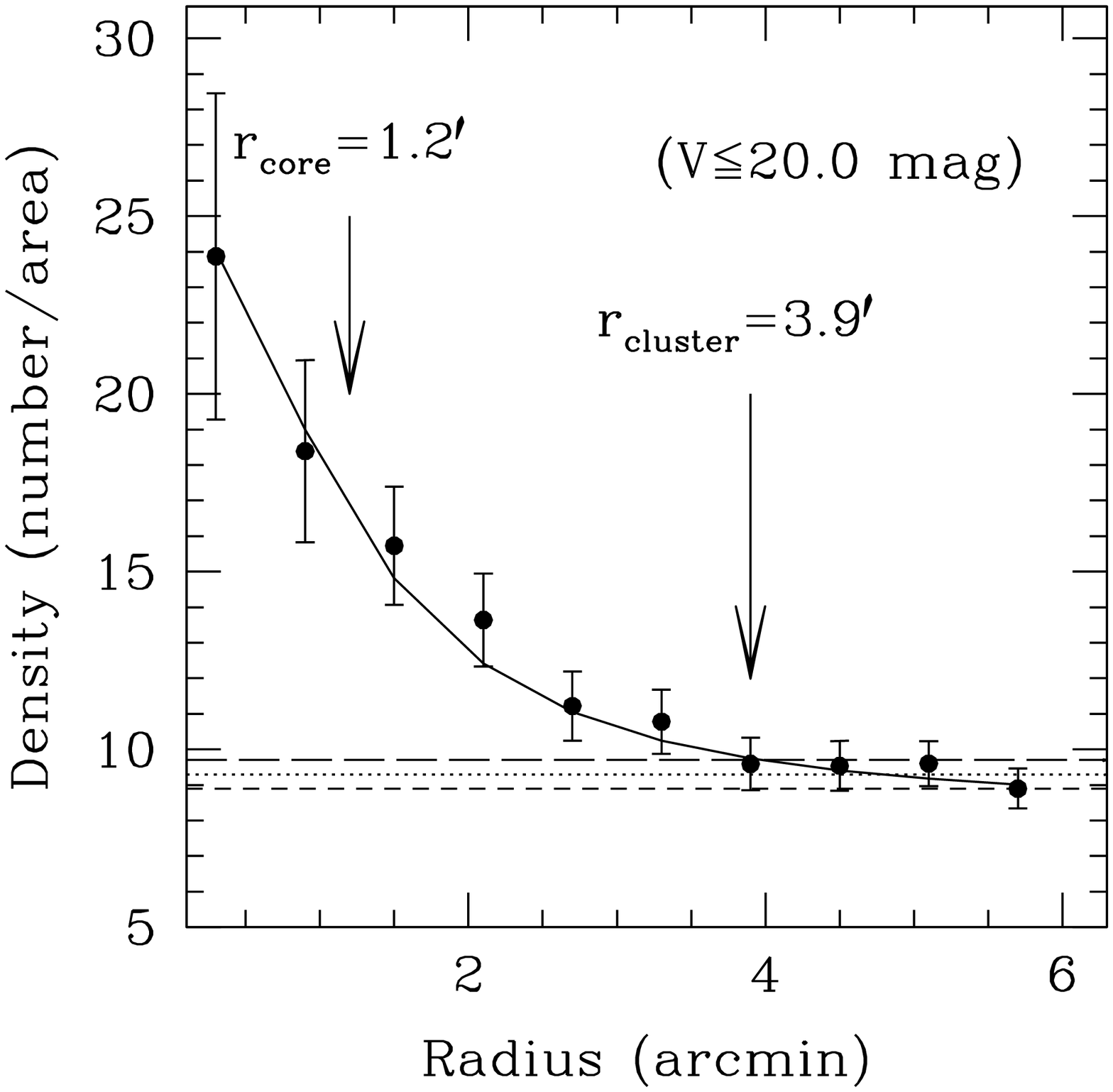}
\vspace{-0.5cm}\caption{Surface density distribution of the cluster SAI 35 in $V$ band. Errors are determined from sampling
 statistics (=$\frac{1}{\sqrt{N}}$ where $N$ is the number of cluster members used in the density estimation at that
point). The smooth line represent the fitted profile of King (1962)  whereas dotted line shows the background density
level. Long and short dash lines represent the errors in background density.}
\label{dens}
\end{center}
\end{figure}

To obtain the structural parameters of SAI 35, we plotted the radial density profile as shown in Fig.~\ref{dens}. We divided
the cluster's region into many concentric rings and number density ($R_{i}$) calculated in each ring by using the formula
$R_{i}$ = $\frac{N_{i}}{A_{i}}$, where $N_{i}$ is the number of stars and $A_{i}$ is the area of the $i^{th}$ zone.
This RDP flattens at $r\sim$ 3.9 arcmin and begins to merge with the background density as shown in the right panel
of Fig.~\ref{dens}. Therefore, we consider 3.9 arcmin as the cluster radius. Our derived value of radius is slightly higher than
3.5 arcmin as obtained by Dias et al. (2014). A smooth continuous line represents fitted King (1962) profile:\\

~~~~~~~~~~~~~~~~~~$f(r) = f_{bg}+\frac{f_{0}}{1+(r/r_{c})^2}$\\

where $r_{c}$, $f_{0}$, and $f_{bg}$ are the core radius, central density, and the background density level, respectively.
By fitting the King model to the RDP of SAI 35, we obtained the values of central density, background density, and core radius
as $25.02\pm2.5$ stars per arcmin$^{2}$, $9.30\pm1.5$ stars per arcmin$^{2}$ and $1.20\pm0.3$ arcmin, respectively.
The background density level with errors is also shown by the dotted lines. The cluster limiting radius, $r_{lim}$, is
calculated by using the formula given by Bukowiecki et al. (2011). The estimated value of the limiting radius is found to be
7.6 arcmin. The concentration parameter is found as 0.8 using the formula given by Peterson \& King (1975).
Maciejewski \& Niedzielski (2007) reported that $R_{lim}$ may vary for individual clusters from 2$R_{c}$ to 7$R_{c}$.
We found that the value of $R_{lim}$ ($\sim$ 6$R_{c}$) for SAI 35 is within the given limit by Maciejewski \& Niedzielski (2007).

\section{Colour-Colour Diagrams}

\subsection{Reddening law}

To understand the nature of extinction law and to find the value of interstellar reddening, we used various
color-color diagrams (CCDs) for SAI 35.

\subsubsection{Total-to-selective extinction value}

Total-to-selective extinction can be
different for cluster members and foreground stars if the size of dust is not the same in the cluster's area and
the interstellar medium along the same line of sight (Mathis 1990). The emitted photons from cluster members are
scattered and absorbed in the interstellar medium by dust particles. The normal reddening law is not applicable in
the line-of-sight that passes through dust, gas and molecular clouds (Sneden et al. (1978)).

\begin{figure}
    \centering
    \includegraphics[width=7.5cm, height=7.5cm]{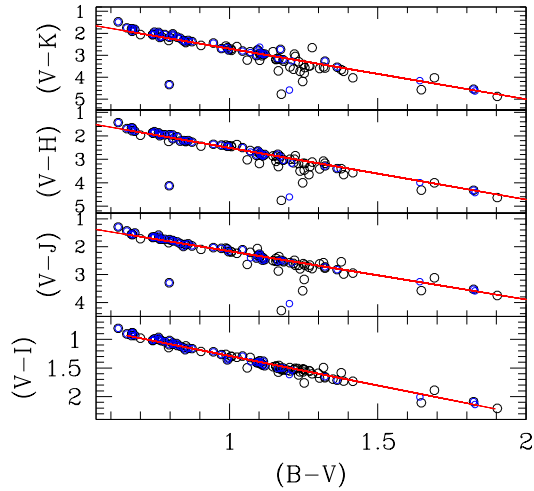}
\vspace{-0.5cm}\caption{The $(\lambda-V)/(B-V)$ CCD for the stars within cluster extent of SAI 35. The continuous 
lines represent the slope determined through the least-squares linear fit.}
  \label{cc2}
  \end{figure}

Chini \& Wargue (1990) suggested $(V-\lambda)/(B-V)$ CCDs to understand the nature of reddening law in which
$\lambda$ is any filter, other than $V$. We plotted various two-color diagrams for SAI 35 as shown in Fig \ref{cc2} to understand the
reddening law. Our obtained values of color-excesses with normal values have been listed in Table \ref{cc_slope}.
The estimated values of color-excesses are in good agreement with the normal values. Since, the stellar color values are
found to be linearly dependent on each other, then a linear equation is applied to calculate the slope $(m_{cluster})$ of
each CCD. Total to selective extinction has been estimated using the relation provided by Neckel \& Chini (1981):\\

\begin{table*}
   \centering
   \caption{A comparison of the extinction law in the direction of the cluster SAI 35 with a normal extinction law given by Cardelli et al. (1989).}
   \begin{tabular}{ccccc}
   \hline\hline
   SAI 35   & $(V-I)/(B-V)$ &  $(V-J)/(B-V)$ &  $(V-H)/(B-V)$  & $(V-K)/(B-V)$ \\
  \hline
   Our derived ratios & $1.36\pm0.2$ & $2.15\pm0.23$ & $2.48\pm0.09$ & $2.70\pm0.11$ \\
   Normal ratios & 1.60 & 2.22 & 2.55  & 2.74  \\
\hline
  \end{tabular}
  \label{cc_slope}
  \end{table*}

$R_{cluster}=\frac{m_{cluster}}{m_{normal}}\times R_{normal}$\\

where $m_{cluster}$ is the normal slope value in each CCD and $R_{normal}$(3.1) is the normal value of total-to-selective
extinction ratio. We have estimated $R_{cluster}$ in different passbands as $\sim$2.9 to 3.3, which is close to the normal
value. So, we found normal reddening law towards the cluster region of SAI 35.

\subsubsection{$(U-B)$ versus $(B-V)$ diagram}

\begin{figure}
    \centering
   \includegraphics[width=7.5cm, height=7.5cm]{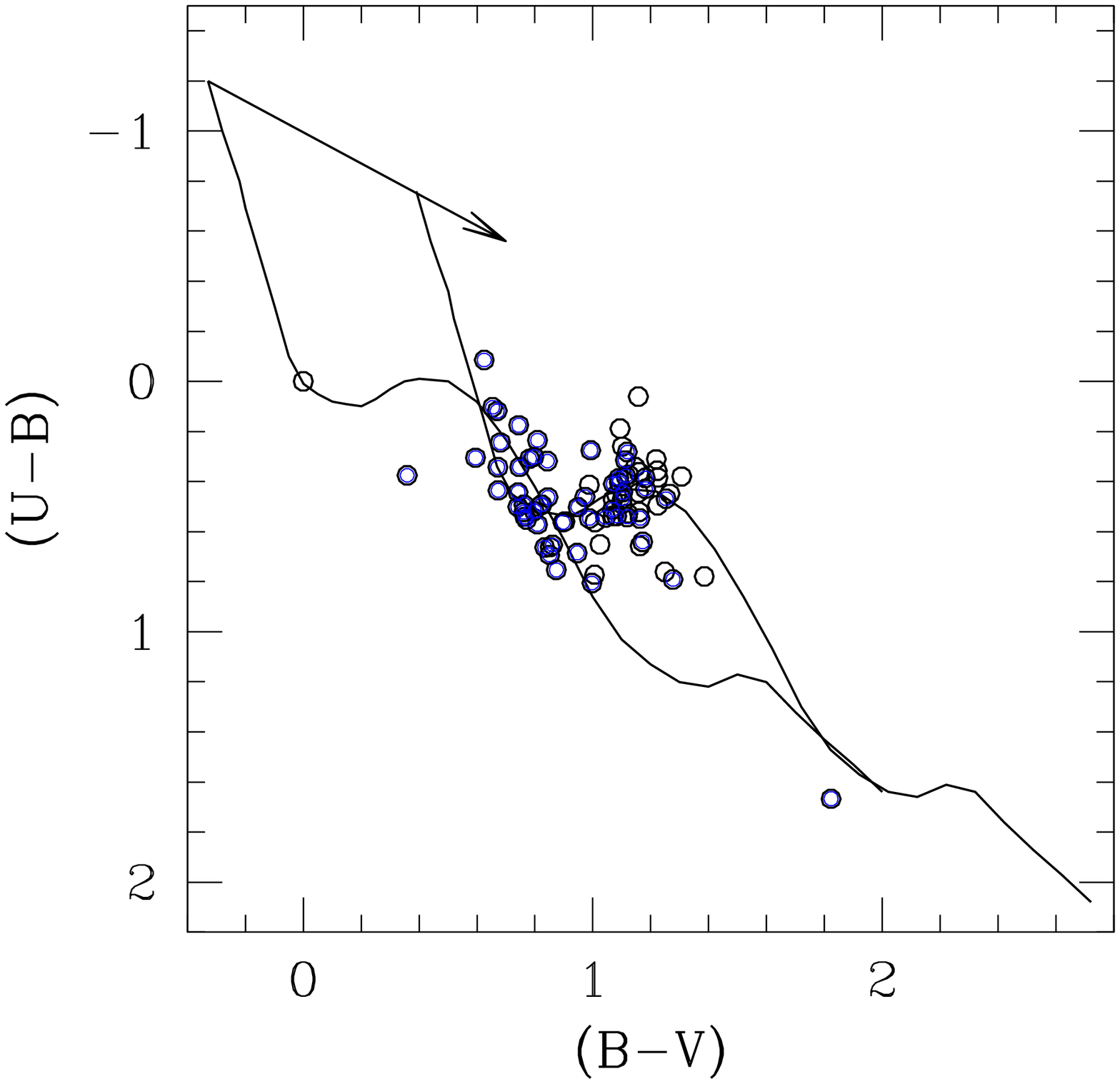}
\vspace{-0.5cm}\caption{The $(U-B)$ versus $(B-V)$ color-color diagram. The continuous curve represents locus of Schmidt-Kaler's
(1982) ZAMS for solar metallicity. Arrow indicates the reddening vector.}
  \label{cc}
  \end{figure}

The knowledge of interstellar reddening is important to obtain main fundamental parameters (age, distance, etc.)
of clusters. In the absence of spectroscopic observations, we can use $(B-V), (U-B)$ color-color diagram for the reddening
estimation of clusters (cf. Becker \& Stock 1954). The resultant $(U-B)$ versus $(B-V)$ plot for SAI 35 is shown in
Fig \ref{cc} using cluster members with membership probability higher than $50\%$. Blue dots are the matched stars with
the catalog provided by Cantat-Gaudin et al. (2018). We have taken the intrinsic zero-age main-sequence (ZAMS) from Schmid-kaler (1982).
The ZAMS is fitted by the continuous curve considering the slope of reddening $E(U-B)/E(B-V)$ as 0.72. By fitting ZAMS to the MS,
we have calculated mean value of $E(B-V)=0.72\pm0.05$ mag for SAI 35. Present estimate of reddening is close to the value 
derived by Kharchenko et al. (2012, 2013) and Sampedro et al. (2017).

   \begin{figure}
     \centering
     \includegraphics[width=7.5cm, height=7.5cm]{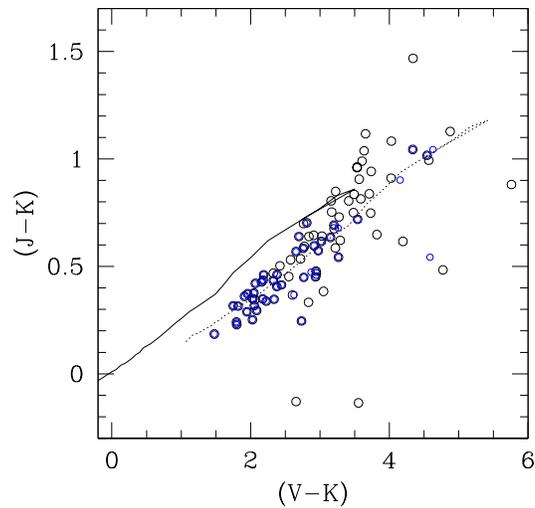}
 \vspace{-0.5cm}\caption{The plot of $(J-K)$  versus $(V-K)$ color-color diagram. Solid and dotted lines are the ZAMS taken
 from Caldwell et al. (1993).}
   \label{cc_jk}
   \end{figure}
\subsubsection{Interstellar extinction in near-IR}

The near-IR photometry is very helpful to understand interstellar extinction (Tapia et al. 1988). Here we have used $JHK$
photometry from the 2MASS database to study the interstellar extinction law. The $(J-K)$ versus $(V-K)$ diagram for SAI 35 is shown
in Fig. \ref{cc_jk}. The ZAMS of solar metallicity is taken from Caldwell et al. (1993) as shown by the solid line. The fit of ZAMS
provides $E(J-K) = 0.33\pm0.04$ mag and $E(V-K) = 1.92\pm0.03$ mag. The ratio $\frac{E(J-K)}{E(V-K)} \sim 0.17\pm0.05$ is in
good agreement with the normal interstellar extinction value of 0.19 is given by Cardelli (1989).

\subsection{Age and distance}

\begin{figure}
\begin{center}
\includegraphics[width=7.5cm, height=7.5cm]{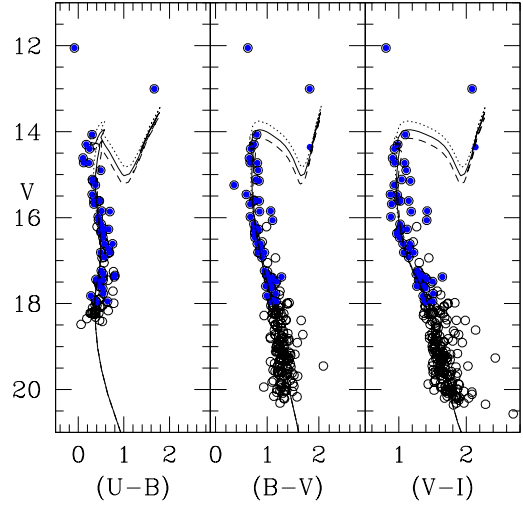}
\vspace{-0.5cm}\caption{ The color-magnitude diagram of the clusters under study. The curves are the
isochrones of (log(age) $=$  8.50 ,8.55 and 8.60). These ishochrones are taken from Marigo et al. (2017). Black
circles are the probable cluster members while the blue circles represent the matched stars with Cantat-Gaudin et al. (2018).}
\label{cmd_gaia_optical}
\end{center}
\end{figure}

\begin{figure}
\begin{center}
\includegraphics[width=7.5cm, height=7.5cm]{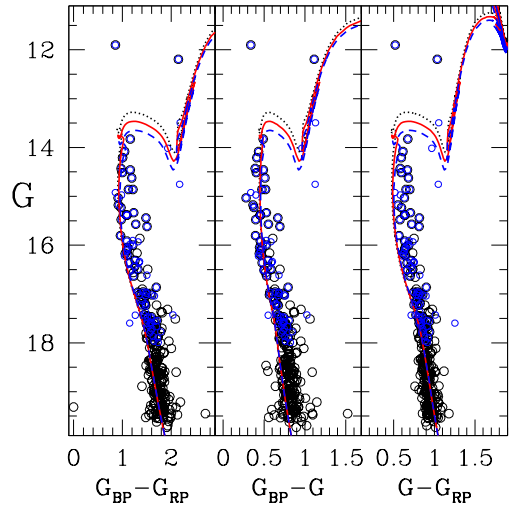}
\vspace{-0.5cm}\caption{ Same as Fig \ref{cmd_gaia_optical} using Gaia EDR3 photometric magnitudes.}
\label{cmd_gaia_optical1}
\end{center}
\end{figure}

Age and distance of OCs are important parameters to trace the galactic structure and to understand the
chemical evolution of Galaxy (Friel \& janes 1993). The main fundamental parameters (reddening, metallicity, distance modulus, age,
etc.) of a cluster can be obtained by fitting the theoretical isochrones to our observed CMDs. We used the theoretical isochrones given
by Marigo et al. (2017) for $Z=0.019$. The  $V/(U-B), V/(B-V)$, $V/(V-I)$, $G/(G_{BP}-G)$, $G/(G_{BP}-G_{RP})$ and $G/(G-G_{RP})$ CMDs
along with visually fitted isochrones are shown in Fig. \ref{cmd_gaia_optical} and  Fig. \ref{cmd_gaia_optical1}. 

We superimpose theoretical isochrones of different ages 
(log(age)=8.50, 8.55 and 8.60) in all the plotted CMDs. Based on this, we have
found an age of $360\pm40$ Myr. Our estimated value of age is same as Sampedro et al. (2017). We obtained distance modulus
$(m-M)$ = $14.60\pm0.2$ mag. The estimated distance modulus provides a heliocentric distance as $2.95\pm0.3$ kpc. Present
distance estimate is in good agreement with Cantat-Gaudin et al. (2018). The Galactocentric coordinates are $X_{\odot} = -2.61$ kpc,
$Y_{\odot} = 1.25$ kpc and $Z_{\odot} = -0.17$ kpc. The Galactocentric distance of the cluster was calculated as 11.20 kpc. The
derived Galactocentric coordinates are in good agreement with Cantat-Gaudin et al. (2018).

\begin{figure}
\begin{center}
\includegraphics[width=7.5cm, height=7.5cm]{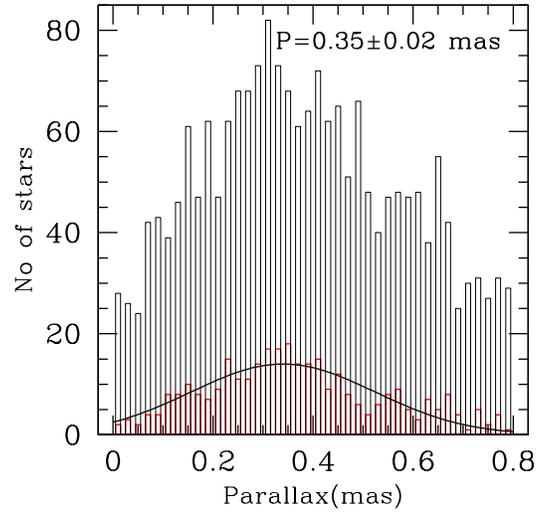}
\caption{Histogram of parallax for SAI 35. Black lines represents all stars in the cluster field whereas the red lines represents
the most probable cluster members. The Gaussian function is fitted to the central bins provides mean value of parallax.}
\label{parallax}
\end{center}
\end{figure}

We have also used the parallax of cluster members to find the distance of SAI 35. The resulting histogram is shown in Fig \ref{parallax}.
The mean parallax is estimated as $0.35\pm0.02$ mas which corresponds to a distance of $2.86\pm0.17$ kpc. The calculated value of parallax
is in good agreement with the value obtained by Cantat-Gaudin et al. (2018). We also estimated distance using the method described by
Bailer-Jones et al. (2018). Thus, we obtained a distance of SAI 35 as $2.90\pm0.15$ kpc. We find a similar value of distance using
the mean parallax and distance modulus of the cluster.

\subsection{Optical and near-IR CMDs}

\begin{figure}
\begin{center}
\includegraphics[width=7.5cm, height=7.5cm]{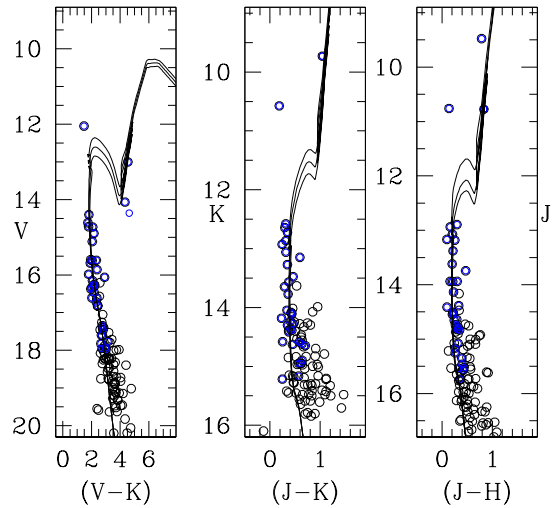}
\caption{Same as Fig \ref{cmd_gaia_optical} of optical and near-IR color-magnitude diagrams.}
\label{cmd_jk}
\end{center}
\end{figure}

Using optical and near-IR data we have re-determined distance and age of SAI 35. We have plotted $V$ versus $(V-K)$, $K$ versus $(J-K)$ and
$J$ versus $(J-H)$ CMDs, which is shown in Fig~\ref{cmd_jk}. The theoretical isochrones given by Marigo et al. (2017) for $Z =$ 0.019
of log(age)=8.50, 8.55 and 8.60 have been over plotted in the CMDs. The apparent distance moduli $(m - M)_{V, (V-K)}$ and $(m-M)_{K, (J-K)}$
turn out to be 14.60$\pm$0.2 and 12.60$\pm$0.3 mag. Using the reddening value estimated in this paper, we have derived a distance of
3.0$\pm$0.3 kpc. Both age and distance determination are in good agreement with the estimates using optical data.

Several fundamental parameters (center, radius, age, distance, reddening etc.) for this object have been derived by several authors
in the literature. Table \ref{comp_para} presents a comparison of our estimated parameters in this paper with previously published
values. All the derived parameters values are comparable with literature.

\begin{table*}
\small
\caption{Comparison of our obtained fundamental parameters for SAI 35 with the literature values}.
\begin{tabular}{lll}
\hline
Parameters & Numerical values & Reference  \\ \hline

(Right ascension, Declination) (deg)    &  (62.70, 46.86)  & Present study\\
                         &  (62.70, 46.87)  & Cantat-Gaudin et al. (2018)\\
                         &  (62.69, 46.86)  & Sampedro et al. (2017)\\

Cluster radius (arcmin)  & 3.9 & Present study\\
                               &3.5 & Dias et al. (2014)\\
$(\mu_{\alpha}cos(\delta), \mu_\delta)$ (mas/yr) &  $(1.10\pm0.01$, $-1.66\pm0.01)$ & Present study\\
                         &   (1.10, -1.64) & Cantat-Gaudin et al. (2018)\\
                         &   (-0.11, -0.90) & Dias et al. (2014)\\
                         &   (-2.36, -6.11) & Kharchenko et al. (2012, 2013)\\
Age (log)        &8.55 &Present study \\
                 &8.55 &Sampedro et al. (2017) \\
                 &8.22 &Kharchenko et al. (2012, 2013) \\
Distance (Kpc)   &    $2.9\pm0.15$ & Present study\\
                 &    2.7          & Cantat-Gaudin et al. (2018)\\
                 &    2.8          & Kharchenko et al. (2012, 2013)\\
$E(B-V)$ (mag)   &    $0.72\pm0.05$   & Present study\\
                 &    0.70           & Sampedro et al. (2017)\\ 
                 &    0.79            & Kharchenko et al. (2012, 2013)\\

 \hline
\end{tabular}
\label{comp_para}
\end{table*}

\section{Dynamical study}

\subsection{Luminosity and mass function}

\begin{figure}
\begin{center}
\includegraphics[width=7.5cm, height=7.5cm]{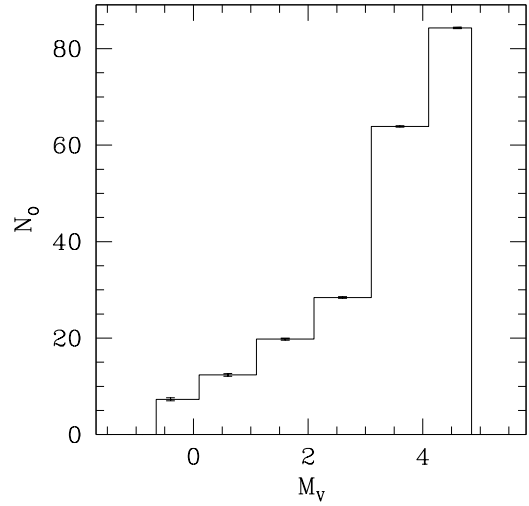}
\vspace{-0.5cm}
\caption{Luminosity function of stars in the region of SAI 35.}
\label{lf}
\end{center}
\end{figure}

Luminosity function (LF) is the distribution of members of a cluster in different magnitude bins. We considered probable cluster members in $V/(V-I)$ CMD
to construct the LF for SAI 35. For the construction of LF, first, we converted the apparent $V$ magnitudes into the absolute
magnitudes by using the distance modulus. Then, we plotted the histogram of LF as shown in Fig \ref{lf}. The interval of 1.0 mag
was picked so that there would be the sufficient number of stars in each bin for statistical usefulness. The LF  of SAI 35 rises
steadily up to $M_{V}$=4.5 mag.

\begin{figure}
\begin{center}
\includegraphics[width=7.5cm, height=7.5cm]{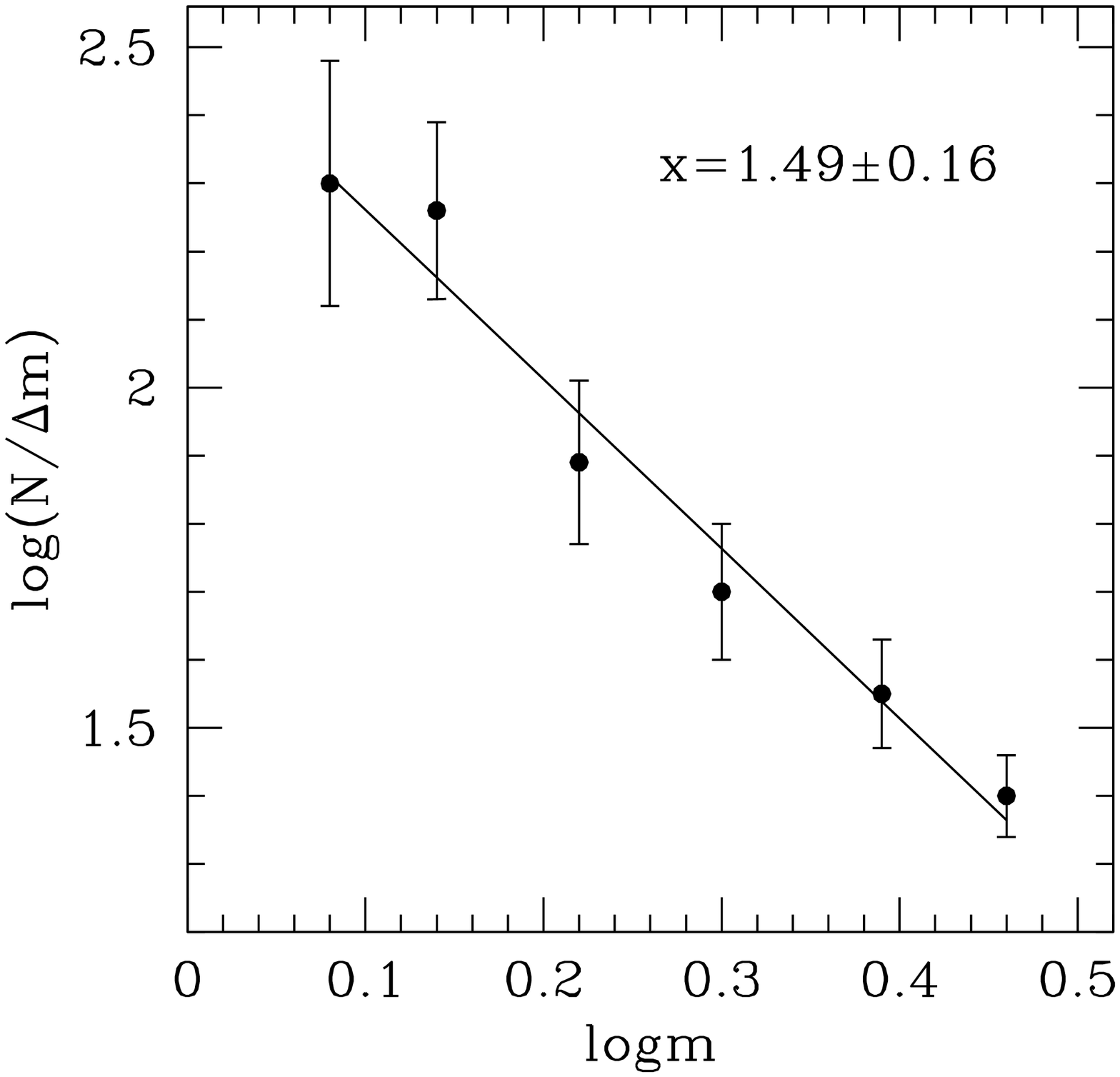}
\vspace{-0.5cm}
\caption{Mass function histogram derived using the most probable members, where solid line indicates the
power law given by Salpeter (1955). The error bars represent $\frac{1}{\sqrt{N}}$.}
\label{mass}
\end{center}
\end{figure}

Mass function (MF) is defined as the distribution of masses of cluster stars per unit volume during the time of star formation.
LF can be converted into the mass function (MF) using a mass-luminosity relation. Since we could not obtain an observational transformation,
so we must depend on theoretical models. To perform the conversion of LF into MF, we used cluster parameters derived in this paper and
theoretical models given by Marigo et al. (2017). The resulting MF is shown in Fig.~\ref{mass}. The mass function slope can be derived
from the linear relation\\

${\rm log}\frac{dN}{dM} = -(1+x) \times {\rm log}(M)+$constant\\

In the above relation, $dN$ represents the number of stars in a mass bin $dM$ with the central mass $M$ and $x$ is the mass function
slope. The Salpeter (1955) value for the mass function slope is $x=1.35$. This form of Salpeter shows that the number of stars in
each mass range decreases rapidly with the increasing mass. 
Our derived MF slope value, $x=1.49\pm0.16$ is in good agreement
with Salpeter's slope within uncertainty. 
Using this value of mass function slope within the mass ranges 1.1~-~3.1 $M_{\odot}$,
total mass was obtained as $\sim$364 $M_{\odot}$.

\subsection{Mass-segregation}

\begin{figure}
\begin{center}
\includegraphics[width=7.5cm, height=7.5cm]{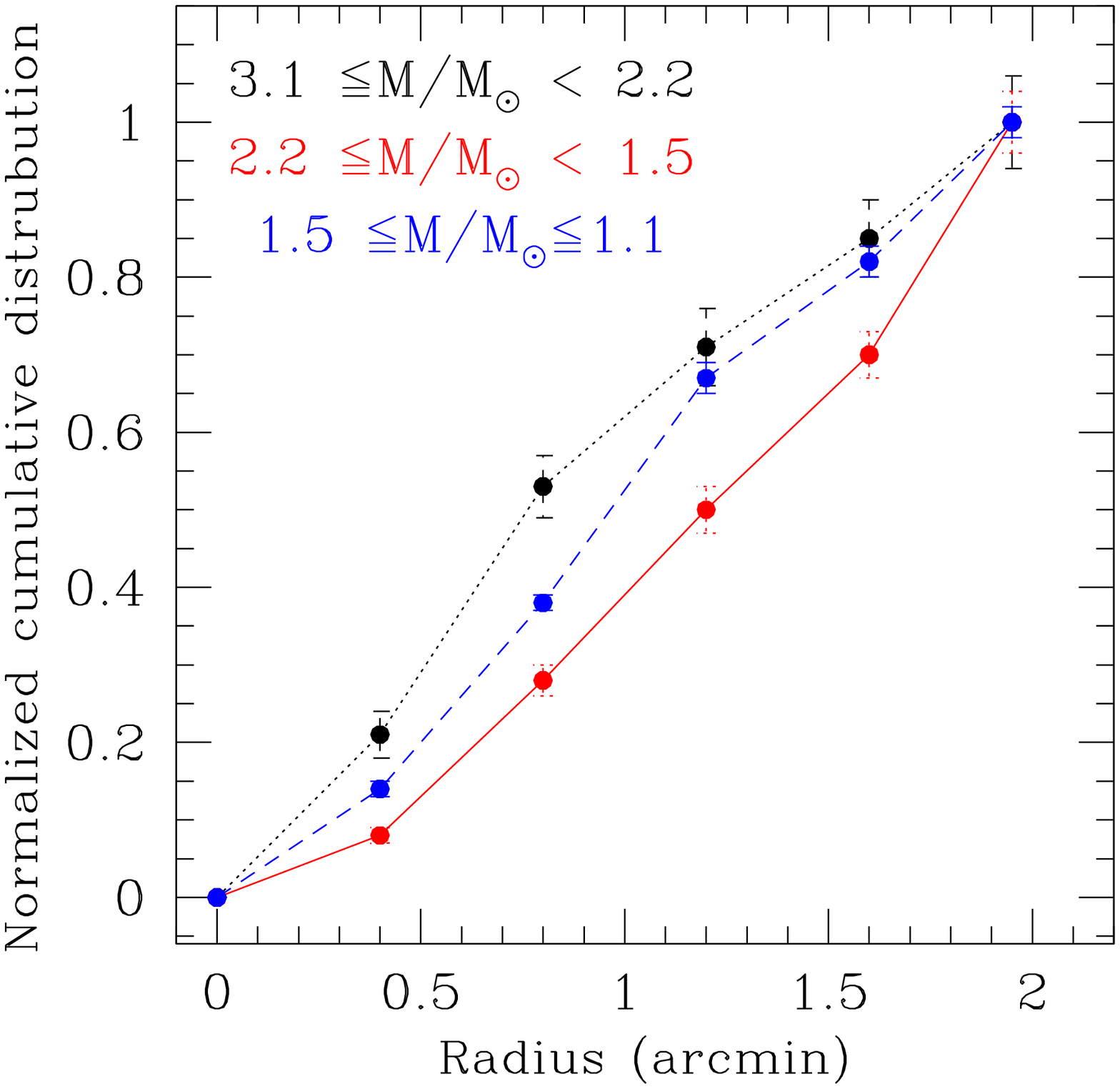}
\vspace{-0.5cm}
\caption{The cumulative radial distribution of stars in various mass range.}
\label{mass_seg}
\end{center}
\end{figure}

As a result of the mass-segregation, massive stars get concentrated towards the center than the fainter stars. Many authors
have reported mass-segregation phenomenon in clusters (Piatti 2016; Zeidler et al. 2017; Dib, Schmeja \& Parker 2018;
Bisht et al. 2020a). To study the effect of mass segregation for the clusters, we plot the cumulative radial stellar
distribution of stars for different masses in Fig~\ref{mass_seg}. This figure exhibits mass-segregation effect
in the clusters under the present study, which means that the massive stars have gradually sunk towards the cluster center as compared to the distribution of the fainter
stars. We have divided the main sequence stars in these three mass ranges,  3.1$\le\frac{M}{M_{\odot}}\le$~2.2,
2.2$\le\frac{M}{M_{\odot}}\le$~1.5 and 1.5$\le\frac{M}{M_{\odot}}\le$~1.1. To check whether these mass distributions
represent the same kind of distribution or not, we have performed Kolmogorov-Smirnov $(K-S)$ test. This test indicates
that the confidence level of mass-segregation effect is 90 $\%$.

Further, it is important to know that the effect of mass-segregation is due to dynamical evolution or imprint of
star formation or both. In the lifetime of star clusters, encounters between its member stars gradually lead to an
increased degree of energy equipartition throughout the clusters. In this process, the higher mass cluster members
accumulate towards the cluster center and transfer their kinetic energy to the more numerous lower-mass stellar component,
thus leading to mass segregation.

\subsection{The Relaxation time}

The time scale in which a cluster will lose all traces of its initial conditions is well represented by its relaxation
time $T_{E}$. The relaxation time is the characteristic time-scale for a cluster to reach some level of energy equipartition.
The relaxation time given by Spitzer \& Hart (1971) stated that:\\

$T_{E} = \frac{8.9 \times 10^{5} N^{1/2} R_{h}^{3/2}}{<m>^{1/2}log(0.4N)}$\\

where $N$ is the number of cluster members, $R_{h}$ is the half-mass radius of the cluster and $<m>$ is mean mass of the cluster
stars. The value of $R_{h}$ was taken as 1.64 pc, which has been assumed as half of the cluster radius derived
by us. Finally, we have estimated the dynamical relaxation time $T_E$ as $\sim$11 Myr. A comparison of cluster age with its relaxation
time indicates that relaxation time is smaller than the age. Therefore, we conclude that SAI 35 is a dynamically relaxed cluster.

\subsection{Tidal radius}

Tidal radius is the distance from cluster center where gravitational acceleration caused by the cluster becomes equal to the
tidal acceleration due to parent Galaxy (von Hoerner 1957). Tidal interactions play important role to understand the
initial structure and dynamical evolution of clusters (Chumak et al. 2010; Dalessandro et al. 2015). The Galactic mass $M_{G}$
inside a Galactocentric radius $R_{G}$ is given by (Genzel \& Townes, 1987),\\

~~~~~~~~~~~~~~~~~~~$M_{G}=2\times10^{8} M_{\odot} (\frac{R_{G}} {30 pc})^{1.2}$\\

Estimated value of Galactic mass inside the Galactocentric radius (see Sec. 4.5) is found as $2.4\times10^{11} M_{\odot}$.
Kim et al. (2000) has introduced the formula for the tidal radius $R_{t}$ of clusters as \\

~~~~~~~~~~~~~~~~~~~~~$R_{t}=(\frac{M_{c}} {2M_{G}})^{1/3}\times R_{G}$\\

where $R_{t}$ and  $M_{c}$ indicate the cluster's tidal radius and total mass (see Sect.~8), respectively. The estimated
value of the tidal radius is $10.22\pm0.82$ pc.

\section{The orbit of the cluster}

We obtained the Galactic orbit of SAI 35 using the Galactic potential models. We adopted the technique described by
Allen \& Santillan (1991) for Galactic potentials. Lately, Bajkova \& Bobylev (2016) and Bobylev et. al (2017) refined
Galactic potential model parameters with the use of new observational data for the galacto-centric distance R $\sim$ 0 to 200 kpc.
The description of the equations used for the Galactic potential models are discussed by Rangwal et al. (2019). We have used
the main fundamental parameters (cluster center ($\alpha$ and $\delta$), mean proper motions ($\mu_{\alpha}cos\delta$, $\mu_{\delta}$),
parallax angles, age and heliocentric distance ($d_{\odot}$)) to obtain orbital parameters of SAI 35. Radial velocity is estimated
as $-91.62\pm6.29$ km/sec from LAMOST DR5 catalog based on membership probability and cluster's VPD.

\begin{table*}
   \centering
   \caption{Position and velocity components in the Galactocentric coordinate system. Here $R$ is the galactocentric
            distance, $Z$ is the vertical distance from the Galactic disc, $U$ $V$ $W$ are the radial tangential and the vertical
            components of velocity respectively and $\phi$ is the position angle relative to the sun's direction.
}
   \begin{tabular}{ccccccccc}
   \hline\hline
   Cluster   & $R$ &  $Z$ &  $U$  & $V$  & $W$ & $\phi$   \\
   & (kpc) & (kpc) & (km/s) &  (km/s) & (km/s) & (radian)    \\
  \hline
   SAI 35 & 10.98 & -0.15 & $60.17 \pm 5.80$  & $-200.86 \pm 2.90$ &  $-6.28 \pm 0.99$ & 0.11    \\
\hline
  \end{tabular}
  \label{inp}
  \end{table*}

We have transformed equatorial velocity components into Galactic-space velocity components ($U,V,W$). The Galactic center is
taken at ($17^{h}45^{m}32^{s}.224, -28^{\circ}56^{\prime}10^{\prime\prime}$) and North-Galactic pole is at ($12^{h}51^{m}26^{s}.282,
27^{\circ}7^{\prime}42^{\prime\prime}.01$) (Reid \& Brunthaler, 2004). To apply a correction for Standard Solar Motion and Motion
of the Local Standard of Rest (LSR), we used position coordinates of Sun as ($8.3,0,0.02$) kpc and its space-velocity components
as ($11.1, 12.24, 7.25$) km/s (Schonrich et al. 2010). Transformed parameters in Galacto-centric coordinate system are listed in
Table \ref{inp}.

\begin{figure*}
\begin{center}
\hbox{
\includegraphics[width=6.2cm, height=6.2cm]{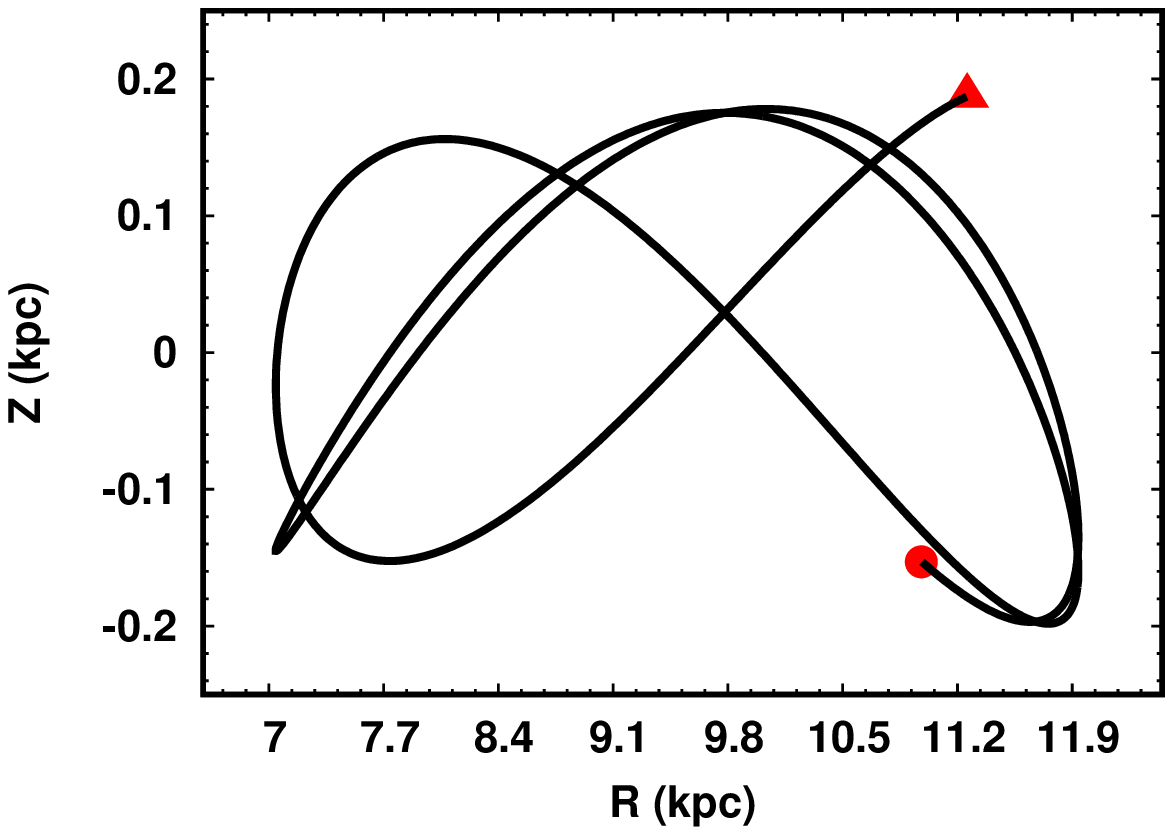}
\includegraphics[width=9.2cm, height=6.2cm]{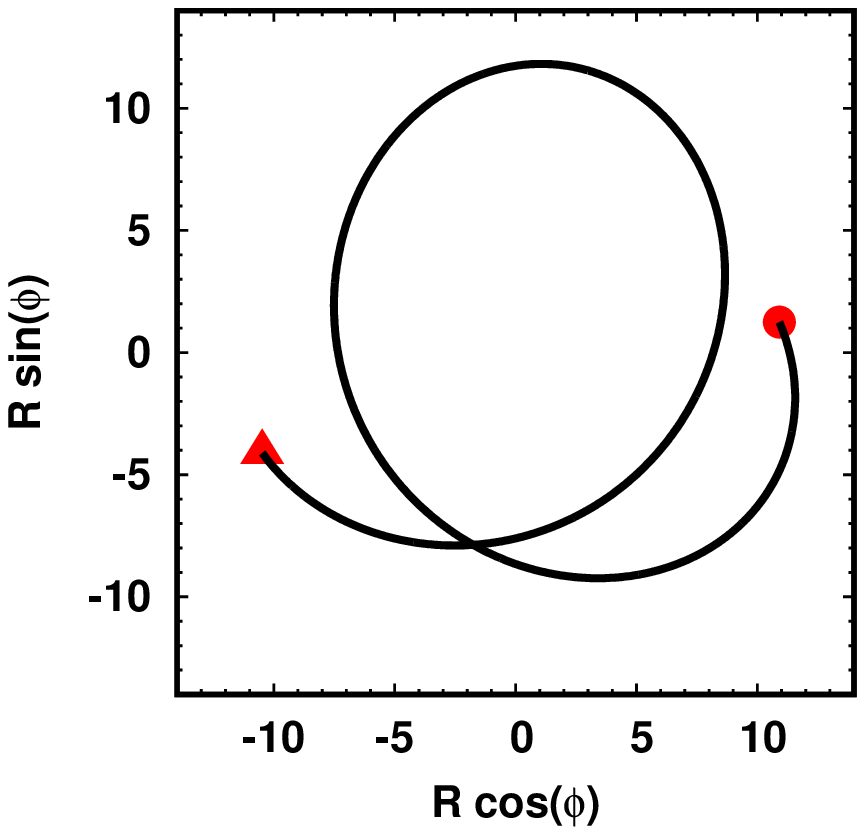}
}
\includegraphics[width=6.2cm, height=6.2cm]{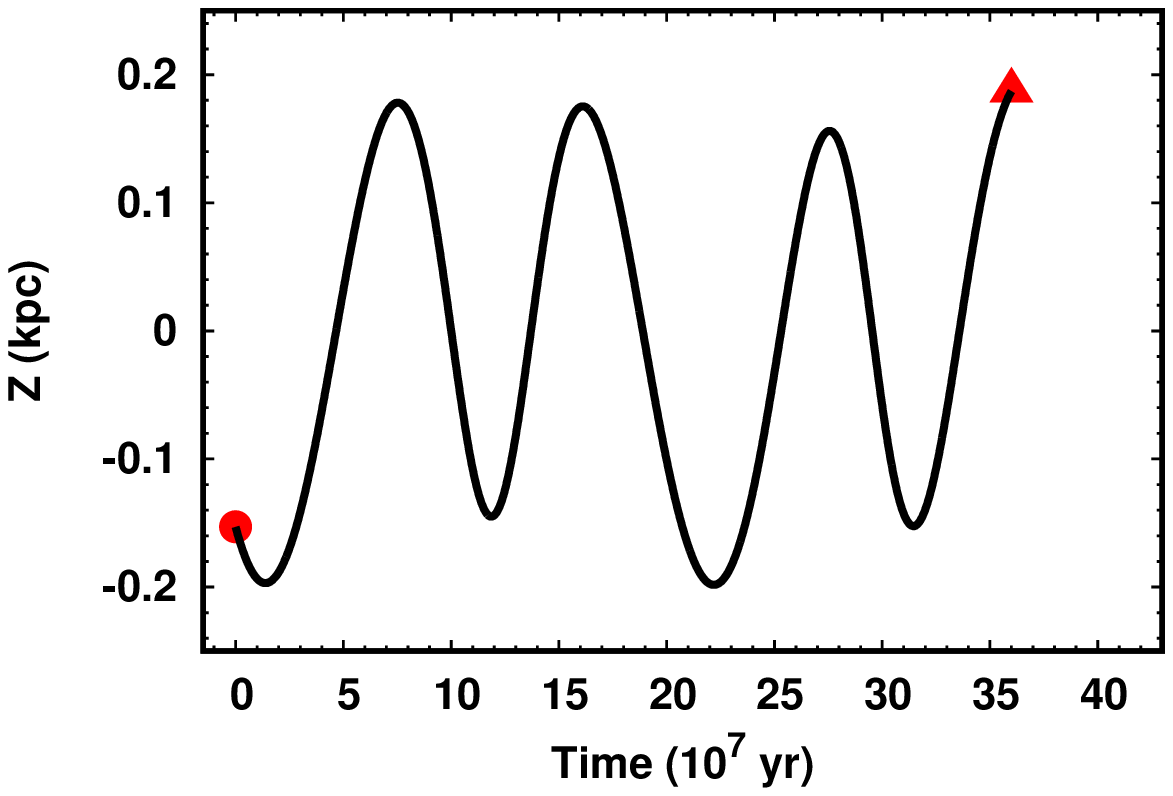}

\caption{Galactic orbits of the cluster SAI 35 estimated with the Galactic potential model described in text in the time interval
of age of cluster. The top left panel shows side view and top right panel shows top view of the orbit. Bottom panel show motion of
SAI 35 in Galactic disc with time. The filled triangle and circle denote birth and present day position of cluster in the Galaxy.}
\label{orbit}
\end{center}
\end{figure*}

\begin{table*}
  \centering
   \caption{Orbital parameters obtained using the Galactic potential model.
   }
   \begin{tabular}{ccccccccc}
   \hline\hline
   Cluster  & $e$  & $R_{a}$  & $R_{p}$ & $Z_{max}$ &  $E$ & $J_{z}$ & $T$   \\
           &    & (kpc) & (kpc) & (kpc) & $(100 km/s)^{2}$ & (100 kpc km/s) & (Myr) \\
   \hline\hline
   SAI 35 & 0.01  & 11.56  & 11.81  & 0.18 & -10.12 & -22.06  & 341 \\
 \hline
  \end{tabular}
  \label{orpara}
  \end{table*}
Fig. \ref{orbit} shows the orbits of the cluster SAI 35. In the top left panel of this figure, the motion of cluster is represented in terms
of distance from Galactic center and Galactic plane, which show a 2-D side view of the orbits. In the top right panel, the cluster motion is
described in terms of $x$ and $y$ components of Galactocentric distance, which shows a top view of orbits. The bottom panel of this
figure indicates that motion of SAI 35 in Galactic disc with time. According to our analysis, SAI 35 follows a boxy pattern. The
value of eccentricity is nearly 0, which demonstrates that the open cluster SAI 35 traces a circular path around the Galactic center.
The birth and present day position of this cluster in the Galaxy are represented by filled circle and triangle in Fig. \ref{orbit}.
The orbit is within the Solar circle. We also calculated various orbital parameters, which are listed in Table \ref{orpara}. Here $e$
is eccentricity, $R_{a}$ is the apogalactic distance, $R_{p}$ is perigalactic distance, $Z_{max}$ is the maximum distance traveled by cluster
from Galactic disc, $E$ is average energy of orbits, $J_{z}$ is $z$ component of angular momentum and $T$ is time period of the
cluster in the orbits. Orbital parameters determined in the present analysis are similar to the parameters determined by Wu et al. (2009).

\section{Conclusions}
\label{con}

We have performed a detailed analysis of the newly discovered open cluster SAI 35 based on Johnson-Cousins UBVI photometry carried out using 1.04-m
Telescope (ARIES, Nainital, India), the 2MASS survey, LAMOST~DR5 catalog and Gaia~EDR3 photometric and astrometric database. We have identified
214 member stars with membership probabilities higher than $50\%$. We investigated the cluster structure, derived the main fundamental
parameters, explained the dynamical study, and determined the galactic orbit of SAI 35. The main outcomes of this study can be
summarized as follows:

\begin{itemize}

\item The new cluster center is estimated as: $\alpha = 62.70\pm0.01$ deg ($4^{h} 10^{m} 48^{s}$) and
      $\delta = 46.86\pm0.004$ deg ($46^{\circ} 51^{\prime} 36^{\prime\prime}$) using cluster members
      based on VPD.

\item Using the radial density profile, the cluster radius is obtained as 
3.9 arcmin (3.3 pc) using the radial density profile.

\item Based on the completeness of CCD data, vector point diagram and membership probability estimation,
      we identified 214 most probable cluster members for SAI 35. The mean PM is estimated as $1.10\pm0.01$ and $-1.66\pm0.01$
      mas yr$^{-1}$ in both the RA and DEC directions respectively.

\item We detected one BSS towards the region of SAI 35 and it was found a confirmed member of the cluster.

\item  From the two color diagram, we have estimated $E(B-V) = 0.72\pm0.05$ mag. We have plotted various CCDs and obtain
       total-to-selective extinction $(R_{V})$ in the range of 2.9 to 3.3. Our Analysis indicates that interstellar
       extinction law is normal towards the direction of SAI 35. From the combined optical and near-infrared data,
       we obtained $E(J-K) = 0.33\pm$0.04 mag while $E(V-K) = 1.92\pm$0.03 mag.

\item The distance of SAI 35 is determined as $2.9\pm0.15$ kpc. This value is well supported by the distance estimated using
      mean parallax of the cluster. Age is determined as $360\pm40$ Myr by comparing the cluster CMD with the solar
      metallicity theoretical isochrones given by Marigo et al. (2017).

\item  The mass function slope is estimated as $1.49\pm0.16$ in the mass range 1.1-3.1 $M_{\odot}$, which is in good
	agreement within uncertainty with the value (1.35) given by Salpeter (1955) for field stars in Solar
       neighborhood. By using this MF slope, we have estimated total mass and mean mass as 364 and 1.70 $M_{\odot}$,
       respectively. 

\item  On the basis of the dynamical evolution study of SAI 35, we found a deficiency of low mass stars in the core. Our study shows a clear
       mass-segregation phenomenon in this cluster. The K-S test indicates $90\%$ confidence level of mass-segregation effect.
       The dynamical relaxation time is estimated as 10 Myr, which is less than the age of the cluster. Our study indicates that SAI 35
       is a dynamically relaxed open cluster.

\item The Galactic orbits and orbital parameters were estimated using Galactic potential models. We found the value of eccentricity $\sim$ 0,
      which concludes that SAI 35 trace circular path around the center of the Galaxy.

\end{itemize}
{\bf ACKNOWLEDGMENTS}\\

The authors thank the anonymous referee for the useful comments that improved the scientific content of the article significantly.
This work has been supported by the Natural Science Foundation of China (NSFC-11590782, NSFC-11421303). Devesh P. Sariya and Ing-Guey Jiang are
supported by the grant from the Ministry of Science and Technology (MOST), Taiwan. The grant numbers are MOST 105-2119-M-007 -029 -MY3
and MOST 106-2112-M-007 -006 -MY3. We also acknowledge Aryabhatta Research Institute of Observational Sciences for great support
during observations. Guoshoujing Telescope (the Large Sky Area Multi-Object Fiber Spectroscopic Telescope LAMOST) is a National Major Scientific
Project build by the Chinese Academy of Sciences. This work has made use of data from the European Space Agency (ESA) mission GAIA
processed by Gaia Data processing and Analysis Consortium (DPAC), (https://www.cosmos.esa.int/web/gaia/dpac/consortium). This publication
has made use of data from the Two Micron All Sky Survey, which is a joint project of the University of Massachusetts and the Infrared
Processing and Analysis Center/California Institute of Technology, funded by the National Aeronautics and Space Administration and
the National Science Foundation. We are also much obliged for the use of the NASA Astrophysics Data System, of the Simbad database
(Centre de Donn$\acute{e}$s Stellaires-Strasbourg, France) and of the WEBDA open cluster database. 

\begin{thebibliography}{}

\bibitem[]{} Ahumada J., Lapasset E., 1995, A\&AS, 109, 375\\\\

\bibitem[]{} Allen, C. \& Santillan, A. 1991, Rev. Mexicana Astron. Astrofis., 22, 255\\\\

\bibitem[]{} Bailer-Jones C. A. L., Rybizki J., Fouesneau M., Mantelet G., Andrae R., 2018, AJ, 156, 58\\\\

\bibitem[]{} Bailyn, C. D., 1995, ARA\&A, 33, 133\\\\

\bibitem[]{} Bajkova, A. T. \& Bobylev, V. V. 2016, Astronomy Letters, 42, 9\\\\

\bibitem[]{}  Balaguer-N\'{u}\~{n}ez L., Tian, K. P., Zhao, J. L., 1998, A\&AS, 133, 387\\\\

\bibitem[]{} Bastian, Nate; Covey, Kevin R., Meyer, Michael R., 2010, ARA\&A, 48, 339B\\\\

\bibitem[]{} Becker W. and Stock J., 1954, Ze. f. Astrophys. 34, 1.\\\\

\bibitem[]{} Binney J., Tremaine S., 2008, Galactic Dynamics. Princeton Univ. press, Princeton, NJ\\\\

\bibitem[]{} Bisht, D., Yadav, R. K. S., Ganesh, S., Durgapal, A. K., Rangwal, G. \& Fynbo, J. P. U. 2019, MNRAS, 482, 1471B\\\\

\bibitem[]{} Bisht, D., Zhu, Q., Yadav, R. K. S. et al., 2020a, MNRAS, 482, 607\\\\

\bibitem[]{} Bisht, D., Elsanhoury W., Zhu, Q., Sariya, D. P. et al., 2020b, AJ, 160, 119\\\\

\bibitem[]{} Bobylev, V. V., Bajkova, A. T. \& Gromov, A. O. 2017, Astronomy Letters, 43, 4\\\\

\bibitem[]{} Bukowiecki, L. et al. 2011, AcA, 61, 231\\\\

\bibitem[]{} Cantat-Gaudin, T., Jordi, C., Vallenari, A., et al. 2018, A\&A, 618A, 93C\\\\

\bibitem[]{} Cantat-Gaudin, T., Anders F., 2020, A\&A, 633, A99\\\\

\bibitem[]{} Caldwell, J. A. R., Cousins, A. W. J., Ahlers, C. C., van Wamelen, P., Maritz, E. J., 1993.
          South African Astron. Observatory (Circ. No. 15).\\\\

\bibitem[]{}  Cardelli J. A., Clayton G. C., Mathis J. S., 1989, ApJ, 345, 245\\\\

\bibitem[]{} Chini R., Wargue W. F., 1990, A\&A, 227, 5\\\\

\bibitem[]{} Chumak Y. O., Platais I., McLaughlin D. E., Rastorguev A. S., Chumak O. V. E., 2010, MNRAS, 402, 1841\\\\

\bibitem[]{} Converse J. M., Stahler S. W., 2011, MNRAS, 410, 2787\\\\

\bibitem[]{} Cui, X. Q., Zhao, Y. H., Chu, Y. Q., et al. 2012, 12, 1197\\\\

\bibitem[]{} Dib S., Scmeja S., Hony S., 2017, MNRAS, 464, 1738\\\\

\bibitem[]{} Dib, Sami., Basu, Shantanu., 2018, A\&A, 614A, 43D\\\\

\bibitem[]{} Dias, W. S., Monteiro H., Caetano T. C. et al. 2014, A\&A, 564A, 79D\\\\

\bibitem[]{} Dalessandro E., et al. 2015, MNRAS, 449, 1811\\\\

\bibitem[]{} Elmegreen, B. G. 2000, ApJ, 539, 342\\\\

\bibitem[]{} Friel E. D., Janes K. A., 1993, A\&A, 267, 75\\\\



\bibitem[]{} Gaia Collaboration et al. 2018a, A\&A, 616, A1\\

\bibitem[]{} Gaia Collaboration et al. 2018b, A\&A, 616, A11\\

\bibitem[]{} Gaia Collaboration. et al. 2020, A\&A, in Prep.\\\\

\bibitem[]{} Genzel R., Townes C. H., 1987, ARA\&A, 25, 377\\\\

\bibitem[]{} Girard, T. M., Grundy, W. M., Lopez, C. E., \& van Altena, W. F. 1989, AJ, 98, 227\\\\

\bibitem[]{} Henden, A., Munari, U. 2014, Contrib. Astron. Obs. Skalnate Pleso, 43, 518\\

\bibitem[]{} Heden, A., Templeton, M., Terrell, D., et al. 2016, VizieR Online Data Catalog, II/336\\

\bibitem[]{} Hur H., Sung, H., Bessel M. S., 2012, AJ, 143, 41\\\\

\bibitem[]{} Johnson H. L., Sandage A. R., 1955, ApJ, 121, 616\\\\

\bibitem[]{} Joshi Y. C., Maurya J., John A. A., Panchal A., Joshi, S., Kumar B., 2020, MNRAS, 492, 3602\\\\

\bibitem[]{} Kaur, Harmeen., Sharma, Saurabh., Dewangan, Lokesh K. et al., 2020, ApJ, 896, 29K\\\\

\bibitem[]{} Khalaj, P., Baumgardt, H., 2013, MNRAS, 434, 3236\\\\

\bibitem[]{} King, I., 1962, AJ 67, 471\\\\

\bibitem[]{} Kim S. S., Figer D. F., Lee, H. M., MOrris M., 2000, ApJ, 545, 301\\\\

\bibitem[]{}  Kumar B., Sagar R., Rautela B. S., Srivastava J. B., Srivastava R. K., 2000, Bull. Astron. Soc. India, 28, 675\\\\

\bibitem[]{} Kharchenko N. V., Piskunov A. E., Schilbach, S., Roeser, S. and Scholz R. D. 2012, A\&A, 543A, 156K\\\\

\bibitem[]{} Kharchenko N. V., Piskunov A. E., Schilbach, S., Roeser, S. and Scholz R. D. 2013, A\&A, 558A, 53K\\\\

\bibitem[]{} Kharchenko N. V., Piskunov A. E., Schilbach, S., Roeser, S. and Scholz R. D. 2016, A\&A, 585A, 101K\\\\

\bibitem[]{} Kronberger, M., Teutsch, P., Alessi, B. et al. 2006, A\&A, 447, 921K\\\\

\bibitem[]{} Kroupa P., 2002, Science, 295, 82\\\\

\bibitem[]{} Landolt A. U., 1992, AJ, 104, 340\\\\

\bibitem[]{} Luo, A. L., Zhang, H. T., Zhao, Y. H., et al. 2012, RAA, 12, 1243\\\\

\bibitem[]{} Maciejewski, G. \& Niedzielski, A., 2007, A\&A, 467, 1065\\\\

\bibitem[]{} Marigo, P. et al. 2017, ApJ, 835, 77\\\\

\bibitem[]{} Mathis, J. S., 1990, ARA\&A, 28, 37\\\\

\bibitem[]{} Neckel T., Chini R., 1981, A\&A, 45, 451\\\\

\bibitem[]{} Peterson, C. J. \& King, I. R., 1975, AJ, 80, 427\\\\

\bibitem[]{} Piatti A. E., 2016, MNRAS, 463, 3476\\\\

\bibitem[]{} Reid M J., Brunthaler A., 2004, ApJ, 616, 872\\\\

\bibitem[]{} Rangwal, G., Yadav, R. K. S., Durgapal, A., Bisht, D. \& Nardiello, D. 2019, MNRAS, 490, 1383\\\\

\bibitem[]{} Sagar R. \& Griffiths W.K., 1998, MNRAS 299, 777\\\\

\bibitem[]{} Sampedro, L., Dias, W. S., Alfaro, E. J., Monteiro, H. and Molino, A. 2017, MNRAS, 470, 3937S\\\\

\bibitem[]{} Sandage A. R., 1953, AJ, 58, 61\\\\

\bibitem[]{} Sandage A. R., 1962, ApJ, 135, 333\\\\

\bibitem[]{} Sariya, D. P., \& Yadav, R. K. S. 2015, A\&A, 584, A59 \\

\bibitem[]{} Sariya, D. P., Jiang, I.-G., \& Yadav R. K. S., 2017, AJ, 153, 134\\\\

\bibitem[]{} Sariya ,D. P., Jiang, I.-G., \& Yadav, R. K. S., 2018, RAA, 18, 126\\\\

\bibitem[]{} Schonrich, Ralph., Binney, James., Dehnen, Walter. 2010, MNRAS, 403, 1829S\\\\

\bibitem[]{} Sneden C., Gehrz R. D., Hackwell J. A., York D. G., Snow T. P., 1978, ApJ, 223, 168\\\\

\bibitem[]{} Salpeter, E. E., 1955, ApJ, 121, 161\\\\

\bibitem[]{} Schmid - Kaler Th., 1982, in Scaitersk., Voigt H. H., eds, Landolt / Bornstein, Numerical Data and Functional
          Relationship in Science and Technology, New series, Group VI, vol. 2b. springer - verlag, Berlin, p. 14\\\\

\bibitem[]{} Skrutskie, M. F., Cutri, R. M., Stiening, R., et al., 2006, AJ, 131, 1163\\\\

\bibitem[]{} Spitzer, L., \& Hart, M. H., 1971, ApJ, 164,399\\\\

\bibitem[]{} Stetson P. B., 1987, PASP, 99, 191\\\\

\bibitem[]{} Stetson P. B., 1992, in Warrall D. M., BiemesderferC., Barnes J., eds, ASp Conf. Ser. Vol. 25, Astronomical
          Data Analysis Software and System I. Astron. Soc. Pac., San Francisco, p. 297\\\\

\bibitem[]{} Soubiran ,C., Cantat-Gaudin, T., et al. 2018, A\&A, 619\\\\

\bibitem[]{} Tapia M., Roth M., Marraco H., Ruiz M. T., 1988, MNRAS, 232, 661\\\\

\bibitem[]{} von Hoerner S., 1957, ApJ, 125, 451\\\\

\bibitem[]{} Wilkinson M. I., Evans N. W., 1999, MNRAS, 310, 645\\\\

\bibitem[]{} Wu, Z. Y., Zhou, X., Ma, J. \& Du, C. H. 2009, MNRAS, 399, 2146\\\\

\bibitem[]{} Yadav, R. K. S. \& Sagar, R. 2002, MNRAS, 337, 133\\\\

\bibitem[]{} Yadav R. K. S., Sariya D. P., Sagar R., 2013, MNRAS, 430, 3350\\\\

\bibitem[]{} Zeidler P., Nota A., Grebel E. K., Sabbi E., Pasquali A., Tosi M., Christian, C. K., 2017, AJ, 153, 122\\\\

\bibitem[]{} Zhao, G., Zhao Y. H., Chu, Y. Q., et al. 2012, RAA, 12, 723F\\\\

\end{thebibliography}
\end{document}